\documentclass{article}

\usepackage{arxiv}

\usepackage{textpos}
\usepackage[numbers]{natbib} 
\usepackage{tabularx,longtable,multirow,subfigure,caption}
\usepackage{fancyhdr} 
\usepackage{url} 
\usepackage[english]{babel}
\usepackage[linesnumbered]{algorithm2e}
\usepackage{amsmath}
\usepackage{amsfonts} 
\usepackage{graphicx}
\usepackage{dsfont}
\usepackage{epstopdf} 
\usepackage{backref} 
\usepackage{array}
\usepackage{latexsym}
\usepackage{hyperref}
\usepackage{titlesec}
\usepackage{booktabs}
\titleformat{\chapter}[display]  
{\normalfont\huge\bfseries}{\chaptertitlename\ \thechapter}{20pt}{\LARGE}  
\titlespacing{\chapter}{0pt}{0pt}{0pt} 

\hypersetup{pdftitle={},
  pdfsubject={}, 
  pdfauthor={},
  pdfkeywords={}, 
  pdfstartview=FitH,
  pdfpagemode={UseOutlines},
  bookmarksnumbered=true, bookmarksopen=true, colorlinks,
    citecolor=black,%
    filecolor=black,%
    linkcolor=black,%
    urlcolor=black}

\newcommand{\probP}{\text{I\kern-0.15em P}}

\date{13 September 2023}

\title{Combining Deep Learning on Order Books with Reinforcement Learning for Profitable Trading}

\author{ \href{https://orcid.org/0009-0005-6158-8068}{\includegraphics[scale=0.06]{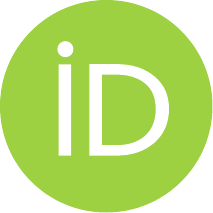}\hspace{1mm}Koti S.~Jaddu} \\
	Department of Computing\\
	Imperial College London\\
	South Kensington, London SW7 2BX \\
	\texttt{koti.jaddu22@imperial.ac.uk} \\
	\And
	\href{https://orcid.org/0000-0001-6846-6649}{\includegraphics[scale=0.06]{orcid.pdf}\hspace{1mm}Paul A.~Bilokon} \\
	Department of Mathematics\\
	Imperial College London\\
	South Kensington, London SW7 2BX \\
	\texttt{paul.bilokon@imperial.ac.uk} \\
}

\begin{document}

\maketitle

\title{Combining Deep Learning on Order Books with RL for Profitable Trading}
\begin{abstract}

High-frequency trading is prevalent, where automated decisions must be made quickly to take advantage of price imbalances and patterns in price action that forecast near-future movements. While many algorithms have been explored and tested, analytical methods fail to harness the whole nature of the market environment by focusing on a limited domain. With the evergrowing machine learning field, many large-scale end-to-end studies on raw data have been successfully employed to increase the domain scope for profitable trading but are very difficult to replicate. Combining deep learning on the order books with reinforcement learning is one way of breaking down large-scale end-to-end learning into more manageable and lightweight components for reproducibility, suitable for retail trading.\\

The following work focuses on forecasting returns across multiple horizons using order flow imbalance and training three temporal-difference learning models for five financial instruments to provide trading signals. The instruments used are two foreign exchange pairs (GBPUSD and EURUSD), two indices (DE40 and FTSE100), and one commodity (XAUUSD). The performances of these 15 agents are evaluated through backtesting simulation, and successful models proceed through to forward testing on a retail trading platform. The results prove potential but require further minimal modifications for consistently profitable trading to fully handle retail trading costs, slippage, and spread fluctuation.

\end{abstract}

\section{Introduction}

In 1992, an internet revolution \cite[Chapter~1, Page~3]{online_trading} disrupted the financial trading industry when the first online brokerage service provider was launched, E*Trade. This quickly replaced traditional trading over the telephone due to its convenience and faster execution. Naturally, the skill ceiling rose as technology improved. Automated algorithmic trading at high speeds, High-Frequency Trading (HFT), was introduced and became popular in the mid-2000s. This trading method involves a bot constantly identifying tiny price imbalances and entering trades before valuations rapidly corrected themselves, ultimately accumulating small profits over time. In 2020, HFT represented 50\% of the trading volume in US equity markets and between 24\% and 43\% of the trading volume in European equity markets while representing 58\% to 76\% of orders \cite[Page~1]{hft_vol}. Although the statistics reveal that HFT is very popular, it hides the fierce competition and immense difficulty- one cannot be perfect in this field. Someone is considered ahead if they find more promising opportunities sooner than their competitors, but these working strategies will not last forever. It is only temporary until a successor arrives, and soon, many will come to surpass. To stay relevant, you must always be the successor. Hence, the emphasis on quantitative research in financial institutions is extensive and in high demand. A portion of such research aims to identify profitable trading strategies that spot opportunities, enter trades, and manage those trades under a millisecond. Do these strategies exist?\\

HFT started with hard-coded rules that overlooked the complex nature of financial markets. A justifiable urge to apply Deep Learning to HFT was later birthed, and as hardware improved along with an abundance of data, so did its potential. Lahmiri et al. \cite{hft_deep_learning} showcase Deep Learning accurately forecasting Bitcoin's high-frequency price data. Kolm et al. \cite{kolm} display remarkable results using Deep Learning to correctly predict high-frequency returns (\textit{alpha}) at multiple horizons for 115 stocks traded on Nasdaq. However, only predicting returns is unlikely to help a desk trader because each trading signal's validity would elapse before the trader could input their order. Reinforcement learning can be the key to interpreting this forecasting to execute high-frequency trades because it can learn an optimal strategy when given return predictions at multiple horizons. Independently, Bertermann \cite{bertermann} has trained a few profitable deep reinforcement learning agents using long, mid, and short-term mean-reverting signals as features. The following work combines Bertermann's reinforcement learning with Kolm et al.'s alpha extraction and investigates its potential in a realistic retail trading setting.\\

Section 1 describes the background and relevant literature to understand the problem while exploring different ideas. The fundamentals of supervised learning will be covered, including their typical pipeline, feed-forward networks, and many network architectures such as Recurrent Neural Networks, Convolutional Neural Networks, and Long Short-Term Memory. Next, reinforcement learning agents such as Q Learning, Deep Q Networks, and Double Deep Q Networks will be touched upon. Limit Order Markets will also be explained to provide more context of the problem. Furthermore, relevant literature will be reviewed which describes different order book feature extraction methods, recent works on using supervised learning and reinforcement learning on the order books, and finally a list of popular performance metrics used to evaluate trading agents. Concepts found in the related work that might not be used in the investigation are also covered in the background for completion and keeping the viewer up to speed.\\

Section 2 discusses the design and implementation of the solutions, which first describes the data collection pipeline and evaluates the quality of the collected data. Next, the design of the supervised learning and reinforcement learning components (including the three agents) are covered while making certain modifications as recommended in the related work. Finally, the section ends with the testing methodology of the models using backtesting and forward testing.\\

Section 3 covers the optimisation strategy for tuning the hyperparameters within the supervised learning and reinforcement learning components. All the parameters are explained here, and reasons for setting certain parameters' values without tuning them are justified.\\

Section 4 evaluates all 15 models after setting their best parameter values using the methodology proposed in Section 4. The performance of the supervised learning and reinforcement learning models are investigated independently through backtesting to support the claims and results observed by their original designers, although modifications were made to try to improve them. This Section also covers the evaluation after combining both models and compares them to a random agent benchmark using statistical testing. The best models were taken through to forward testing and the results are presented. Finally, the agents are explained using heatmaps to investigate what values of input lead to buying and selling behaviours.\\

Section 5 concludes the findings showing potential and highlights the limitations of this work. Further improvements are proposed to have these algorithms overcome the difficulty of submitting high-frequency trades profitably at the retail level.

\section{Background and Related Work}
This Section covers the theory required to further understand the problem at hand and break it apart into its components: supervised learning, reinforcement learning, and limit order markets. Relevant published work will also be reviewed which will be built upon.

\subsection{Supervised Learning}

Machine Learning is a growing field and has been popular in finding statistical patterns in noisy data using reusable blocks in its robust pipeline. This Section introduces the most popular paradigm in Machine Learning and touches on a few architectures of these blocks that will be mentioned in this work. Please refer to \cite{badillo, johnson, dolphin} for more detail.\\

There are three Machine Learning paradigms: supervised learning, unsupervised learning, and reinforcement learning. Unsupervised learning is irrelevant to this work, so it will be ignored. Supervised learning is the study of using existing labelled data to automate learning quantitative relationships between its inputs (\textit{features}) to predict the labels of unlabelled data. The terms labels and classes will be used interchangeably, which are one or more values/text stamped to each record of features. Labels can be continuous (\textit{regression}) or discrete/categorical (\textit{classification}).

\subsubsection{Pipeline}

The pipeline consists of first splitting existing data into a training and held-out testing set (usually of ratio 4:1) to evaluate the model. A model is instantiated, and the data in the training set are fed as batches to update the model- the training process. The training set is separated into its features and labels. The features are passed through the model, and a prediction is made. This prediction is compared with the actual label, and the model is updated through a process called \textit{back-propagation} (please refer to \cite{kostadinov}). This encourages the model to make correct predictions when similar features are observed. When the entire data is passed through the model, one \textit{epoch} is completed. Many epochs may be required to train a model. In the end, the held-out test set is used to evaluate the model with unseen data. This is the overall pipeline, but more modifications can improve performance, e.g. splitting the training into a validation set to prevent \textit{overfitting}. This is when the model fits the training data well but generalises poorly and fails to encounter unseen data. Here is a list of \textit{hyper-parameters} that have to be selected before the training process:

\begin{itemize}
    \item Network Architecture: the number of layers, nodes, and what functions to use
    \item Loss Function: measures how far the predictions are from the actual values
    \item Optimiser: the method used to update the parameters of the model
    \item Learning Rate: the rate at which the model should update its parameters
    \item Number of Epochs: the number of times iterating the dataset during training
    \item Batch size: the number of records to expose to the model before each update
\end{itemize}

\subsubsection{Layers in a Network}
A network has an input layer, optional hidden layer(s), and an output layer. Data is passed from the input layer to the output layer, and data is manipulated as it propagates through. Layers are only connected to adjacent layers. In a network, one can choose the type of layers and how many are used. The linear layer is the simplest but widely used. It processes the previous layer's output by multiplying that vector with a weight vector, then adding a bias vector, and finally applying a non-linear activation function to allow the modelling of non-linear relationships. If the layer is the network's first (input) layer, it takes in the features as input. These weights and bias vectors are parameters that are learned by the network during training. Popular activation functions include ReLU, Sigmoid, and TanH.\\

\begin{figure}[!ht]
\centering
\includegraphics[width = 0.6\hsize]{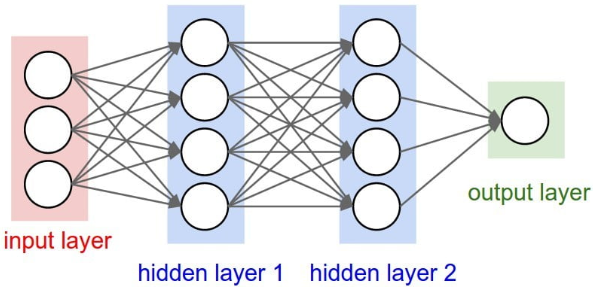}
\caption{Visual representation of a typical linear network. This network has an input layer with three nodes, two hidden layers each with four nodes, and an output layer with one node. Each line represents the passing of a calculated value to a node in the next layer. This graphic was taken from \cite{johnsondeep}.}
\label{fig:network}
\end{figure}

The network designer can also set the number of nodes in each layer, which can be seen as the capacity of the layer. Each node has weighted input and output connections, where the number of connections depends on the number of nodes in adjacent layers. The nodes are usually fully connected but only to nodes in adjacent layers. The computation required to manage an extensive network is demanding, but a compromise is needed to represent more complex functions. Giving a network too much capacity, i.e. too many nodes, can result in overfitting. An illustration of a typical linear network is presented in figure \ref{fig:network}.

Convolutional layers in Convolutional Neural Networks (CNN) have reduced parameters compared to linear layers because nodes are not fully connected to nodes in adjacent layers. On top of this, the weights and bias vectors are shared across all nodes in the layer, allowing for the processing of high-dimensional data like images. Ultimately, convolutional layers are good at detecting themes in local regions and are invariant to translation. Pooling operations down-sample image data as it propagates through the network to consolidate its learned features and spot global patterns.\\

\begin{figure}[h!]
\centering
\includegraphics[width = 0.69\hsize]{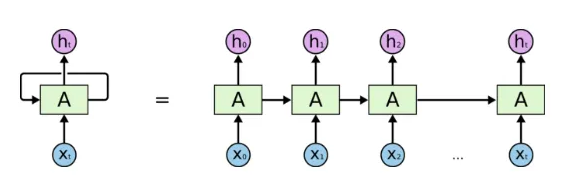}
\caption{Visual representation of an unrolled Recurrent Neural Network, $X$ is the input sequence and $h$ is the output sequence produced. This graphic was taken from \cite{mittal}.}
\label{fig:rnn}
\end{figure}

Recurrent Neural Networks (RNN) are designed to work with sequential data such as natural language and time series data. This is because they can create a sequence of outputs while propagating the result back into itself to combine it with the next token in the input sequence. After the input is passed, the output will then be the input to the next layer, as mentioned before. A graphic is shown in figure \ref{fig:rnn}.

However, there are many disadvantages to using RNNs. The first is that it has short-term memory, meaning it fails to capture dependencies far apart in input sequences. The second is the gradient vanishing and exploding problem. The longer the input sequence, the more times the weight was multiplied by itself to produce an output. A recurring multiplication with a weight smaller than one eventually becomes too small, and a recurring multiplication with a weight larger than one eventually becomes too big. These extremities make it hard for the network to learn efficiently.\\

\begin{figure}[h!]
\centering
\includegraphics[width = 0.7\hsize]{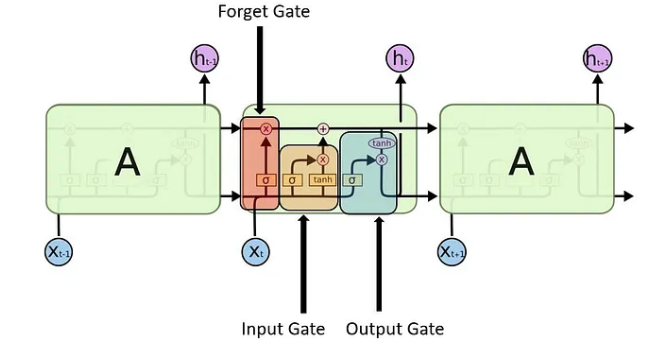}
\caption{Visual representation of an LSTM cell with its three gates, $X$ is the input sequence and $h$ is the output sequence produced. This graphic was taken from \cite{mittal}.}
\label{fig:lstm}
\end{figure}

Long Short-Term Memory (LSTM) networks resolve these obstacles by operating three gates. The \textit{forget gate} decides which parts of the long-term memory state to remove. The \textit{input gate} controls what new information to add to the long-term memory state. The \textit{output gate} finally applies a filter combining long and short-term memory with the inputs to return an output. Please refer to figure \ref{fig:lstm}.

\subsection{Reinforcement Learning}

Reinforcement learning is based on observing rewards from an environment for every action taken and learning behaviours that maximise the reward. The field is particularly popular in Robotics to create agents that operate in the real world. This Section explains the algorithms that will be mentioned in this work. For more information, please refer to \cite{rl_textbook}.

\subsubsection{Terminology}
First, here are some key terms that will be used. An action is either discrete (from a finite action set) or continuous (from a real-valued vector), allowing the agent to interact with the environment. A \textit{state} is a minimal but sufficient representation of the environment. It comprises a fixed length tuple with each value indicating the situation of a feature. The environment can be in many different states during an \textit{episode}. An episode is a simulation involving agent interactions with the environment from the initial state to the end state, and the goal is to learn what actions should be taken in what states to maximise the \textit{reward}- desired reactions from the environment observed after every action. Ultimately, learning involves converging to an optimal \textit{value function} or \textit{policy}. A value function $V(s, a)$ takes input a \textit{state-action pair} (the current state and action) and returns the expected return. A policy $\pi(s)$ takes the current state as input and returns the optimal action.

\subsubsection{Q Learning}
When the dynamics of the environment are not known, \textit{model-free} learning is preferred, which directly estimates the optimal policy or value function. \textit{Monte Carlo} sampling can be used to obtain $(\text{state} \, s, \text{action}\, a, \text{reward}\, r, \text{next state} \,s')$ data which the agent can learn from. \textit{Temporal-difference} (TD) methods combine this sampling with \textit{bootstrapping}- using the estimated values of proceeding states to approximate the value of the current state. Q Learning is a TD method which stores a $Q$ table containing values that estimate the maximum discounted future reward for each action taken at each state. After each reward is observed, these values are updated using the Bellman equation \cite[Page 81]{rl_textbook}. The Q Learning algorithm can be seen in Algorithm \ref{alg:qLearning}. Three other parameters exist. The \textit{learning rate} [$\alpha \in (0, 1)$] affects how much each value is changed after each update. The \textit{discounted future reward factor} [$\gamma \in (0, 1)$] controls how much the values in proceeding states affect the current state. The \textit{exploration rate} [$\epsilon \in (0, 1)$] is the probability of performing a random action to explore more state-action pairs. The highest value across the actions in the $Q$ table for a particular state is the ($\epsilon$\textit{-greedy}) action to take.

\begin{algorithm}
\caption{Q Learning for estimating $\pi \approx \pi_*$ as in \cite[Page 153]{rl_textbook} }\label{alg:qLearning}
Initialise $Q(s, a)$ arbitrarily \\
\Repeat{$n$ episodes completed}{
    Initialise  $s$ \\
    \Repeat{$s$ is terminal}{
        Choose $a$ from $s$ using policy derived from $Q$, \probP(random action) = $\epsilon$\\
        Take action $a$, observe $r$, $s'$\\
        $Q(s, a) \gets Q(s, a) + \alpha [r + \gamma \, \underset{a'}{\max} \, Q(s', a') - Q(s, a)]$\\
        $s \gets s'$\\
    }
}
\end{algorithm}
\subsubsection{Deep Q Learning Network (DQN)}
Q Learning requires storing a table, meaning that a continuous state space will need to be abstracted to buckets. There is no right way to create these buckets; even after that, a state outside the table's coverage could be visited. Deep Q Learning solves this issue by using a neural network to represent the value function instead of a table, which can generalise well in continuous state spaces. A \textit{target network} helps stabilise the learning process by providing a reference for calculating target Q values and updates less frequently compared to the primary network, which continues to change at every backpropagation step. An \textit{experience replay buffer} $D$ allows samples to be reused and improves computational efficiency by training in mini-batches. Algorithm \ref{alg:dqn} shows the Deep Q Learning algorithm.

\begin{algorithm}
\caption{DQN as in \cite[Page 6]{dqn} but with a target network}\label{alg:dqn}
Initialise replay memory $D$\\
Initialise primary $Q(s, a; \theta)$ and target $Q'(s, a; \theta')$ networks\\

\Repeat{$n$ episodes completed}{
    Initialise  $s$ \\
    \Repeat{$s$ is terminal}{
        Choose $a$ from $s$ using policy derived from $Q$, \probP(random action) = $\epsilon$\\
        Take action $a$, observe $r$, $s'$\\
        Store transition $(s, a, r, s')$ in $D$\\
        Sample random mini-batch of transitions $(s_i, a_i, r_i, s_i')$ from $D$\\
        $y_i \gets r_i + \gamma \, \underset{a'}{\max} \, Q'(s_i', a'; \theta')$\\
        Perform a gradient descent step on $(y_i - Q(s_i, a_i; \theta))^2$ wrt $\theta$\\
        $s \gets s'$\\
        Every $C$ steps, set $Q' \gets Q$
    }
}
\end{algorithm}

\subsubsection{Double Deep Q Learning Network (DDQN)}
A DDQN advances from the DQN by addressing the \textit{overestimation bias}, meaning that the learned Q values are higher than they should be due to the maximum operation used on line 10 in Algorithm \ref{alg:dqn}. This leads to sub-optimal decision-making and is mitigated in the DDQN by using the primary network to evaluate the Q-value of the action selected by the target network. This makes it less likely for both networks to overestimate the Q value of the same action.

\begin{algorithm}
Initialise replay memory $D$\\
Initialise primary $Q(s, a; \theta)$ and target $Q'(s, a; \theta')$ networks\\

\Repeat{$n$ episodes completed}{
    Initialise  $s$ \\
    \Repeat{$s$ is terminal}{
        Choose $a$ from $s$ using policy derived from $Q$, \probP(random action) = $\epsilon$\\
        Take action $a$, observe $r$, $s'$\\
        Store transition $(s, a, r, s')$ in $D$\\
        Sample random mini-batch of transitions $(s_i, a_i, r_i, s_i')$ from $D$\\
        $y_i \gets r_i + \gamma \,  Q(s_i', \, \underset{a'}{argmax} \, Q'(s_i', a'; \theta'); \theta)$\\
        Perform a gradient descent step on $(y_i - Q(s_i, a_i; \theta))^2$ wrt $\theta$ \\
        $s \gets s'$ and $\theta' \gets \tau * \theta + ( 1 - \tau) * \theta'$\\
    }
}
\caption{DDQN as explained by Hasselt et al. in 2016 \cite{ddqn}}\label{alg:ddqn}
\end{algorithm}

\subsection{Limit Order Markets}
This Section describes what the prior theoretical knowledge will be applied to and how it operates, as well as existing methods for transforming raw data for deep feature extraction. The following explanation of Limit Order Markets is based on Hambly's slides \cite{intro_to_limit_order_book_markets}, which contains an excellent illustrative introduction.\\

A market is where buyers and sellers meet to exchange goods. A seller will display their asset at a price and wait for a buyer to accept the trade. Contrarily, a buyer can offer a quote and wait for a seller to accept. So how do buyers and sellers make money? You have to make two transactions to start and finish with no inventory (hold no stock). A buyer can make money from buying an asset at price $X$ and then selling it at a higher price $Y$. The profit is $Y - X > 0$. A seller can also make money, but there is an extra complication. They first would need to borrow an asset, sell it at price $X$, re-buy it at a lower price $Y$, and then return the asset to its owner. The profit is $X - Y > 0$. You can therefore make money from buying then selling or selling then buying assets.\\

Once trading was accessible on the Internet, markets needed to operate in an organised and secure fashion. Limit Order Markets are popular for addressing this. It features a \textit{Limit Order Book (LOB)} which keeps track of a queue of \textit{limit orders} at each discrete price bucket (\textit{ticksize}) and are filled by the counterpart in a first-in-first-out (FIFO) manner. A \textit{buy limit order} is a request to buy \textbf{below} the mid-price, and a \textit{sell limit order} is a request to sell \textbf{above} the mid-price. The \textit{volume} is attached to each limit order, which is the number of shares the user wishes to buy or sell.
\begin{figure}[h!]
\centering
\includegraphics[width = \hsize]{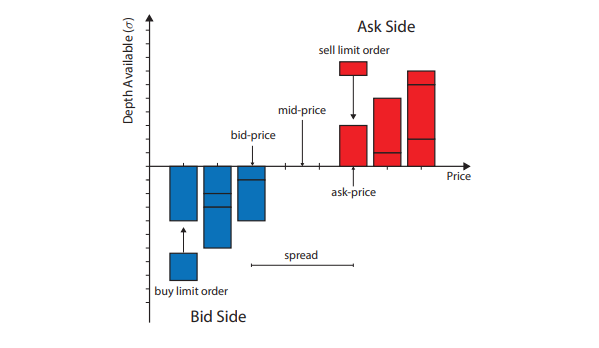}
\caption{Limit Order Book (LOB) illustration as found in \cite[Slide~17]{intro_to_limit_order_book_markets}}
\label{fig:orderbook}
\end{figure}
Figure \ref{fig:orderbook} shows a visual representation of the ask side (all the sell limit orders- red) and the bid side (all the buy limit orders- blue) along with the corresponding aggregated volume of shares at each price level displayed as \textit{depth available}. The \textit{mid-price} represents the current price of the asset and is calculated in the following way:
\begin{equation}
    \textbf{p}_t^{MID} := \frac{a_t + b_t}{2},
    \label{midprice}
\end{equation}
where $a_t$ is the best (lowest) ask price and $b_t$ is the best (highest) bid price at time $t$. The \textit{bid-ask spread} is the difference between the best ask price and the best bid price $a_t - b_t$. Tighter spreads indicate greater \textit{liquidity} (higher trading activity), which is desirable to ensure limit orders are likely to be filled. When new limit orders enter the order book, they are added to the end of the queue at the corresponding price level, as shown in figure \ref{fig:orderbook}. Once an order is sent, it can be cancelled as long it has not been filled.

Limit orders are filled and leave the queue only at the best bid-ask prices. This happens when an incoming counterpart \textit{market order} of \textbf{sufficient} volume is processed; otherwise, the limit order will be partially filled or yet to be filled if it is at the back of the queue. A \textit{market order} is executed immediately, and \textit{market buy orders} are matched with the limit sell orders at the best ask price. In contrast, \textit{market sell orders} are matched with the limit buy orders at the best bid price. Once all limit orders at either the best bid or ask price level are filled, the best bid or ask price moves to the next best bid or ask price level and the mid-price changes accordingly. Similarly, if a new limit order is submitted inside the spread, the mid-price changes because the best bid or ask price has advanced to that new price level. These two scenarios move the market.

\subsection{Literature Review}
This Section uncovers the related work done in the space of combining deep learning on the order book with reinforcement learning. This includes popular order book feature extraction methods employed as well as reporting the methods and results from using supervised learning and reinforcement learning techniques on the order book. The Section concludes with performance metrics for evaluating trading agents.

\subsubsection{Order Book Feature Extraction Methods}
Feature extraction is a crucial step when dealing with complex raw data such as the order book. It involves selecting, transforming, and representing the raw data in a more meaningful way, easing the learning process of any model. This is because feature extraction decreases the dimensionality of the data, attempting to mitigate the curse of dimensionality issue \cite{mario}, which proves an increase in computational complexity when dealing with higher dimensions. However, if the number of dimensions is greatly reduced, then too much useful information is abstracted, so there must be a balance. Feature extraction also reduces the noise in the data as incorporating domain knowledge and expertise allows for keeping what is required for learning and discarding the rest.\\

 As a starting point, here is the raw state of the LOB which will be built upon. It can be abstracted such that each price level has attached a value: the aggregated sum of volumes of limit orders at that price. If we look at the first ten non-empty levels of the order book on each side (bid and ask), the state of the LOB at time $t$ can be written as a vector:

\begin{equation}
    \textbf{s}_t^{LOB} := (a_t^1, v_t^{1,a}, b_t^1, v_t^{1,b}, ..., a_t^{10}, v_t^{10,a}, b_t^{10}, v_t^{10,b})^T \in \mathbb{R}^{40},
    \label{baselob}
\end{equation}
where $a_t^{i}, b_t^i$ are the ask and bid prices at the $i$-th level at time $t$ and $v_t^{i,a}, v_t^{i,b}$ are the respective aggregated sum of volumes of limit orders at the level. There are many features that can be extracted from the LOB, but three popular methods suitable for predicting mid-price jumps will be explained: price, volume, and order flow.

\paragraph{Price Extraction Methods}

From initial speculation, removing the volume elements from equation \ref{baselob} leaves the price components, and so a valid price extraction method could be the following:

\begin{equation}
    \textbf{s}_t^{price} := (a_t^1, b_t^1, ..., a_t^{10}, b_t^{10})^T \in \mathbb{R}^{20},
\end{equation}

Although this halves the number of dimensions, there is room for improvement. A simple but useful feature that can be extracted from the LOB is the mid-price, which infers the state of the best bid and ask prices. The calculation for mid-price has been covered in equation \ref{midprice}, and keeping it consistent with the introduced notation- the most abstracted state of the LOB is the following.

\begin{equation}
    \textbf{s}_t^{mid-price} := \frac{a_t^1 + b_t^1}{2} \in \mathbb{R},
\end{equation}

This reduces the number of dimensions representing the state of the market from 40 if using ten  non-empty bid-ask levels to 1, which is very computationally efficient but removes a lot of information that could be useful. Nevertheless, there exist many trading algorithms that only use the mid-price. Two popular examples include:
\begin{itemize}
    \item Mean reversion strategies \cite{cmc_meanreversion} identify deviations from the mean calculated across a period of historical mid-prices and trade hoping price will correct itself towards the mean if deviated above a threshold.
    \item Momentum strategies \cite{cmc_momentum} identify high volatile movements in mid-price and trade hoping for price to continue in that direction.
\end{itemize}

Such algorithms require recent historical mid-price data so here is a more appropriate feature extraction method that captures the change in mid-price across ten consecutive timesteps:

\begin{equation}
    \textbf{s}_t^{\Delta mid-price} := \left( \frac{(a_i^1 + b_i^1)}{2} - \frac{(a_{i-1}^1 + b_{i-1}^1)}{2} \, \Big| \, i \in [ t-9, t ] \right) \in \mathbb{R}^{10},
\end{equation}

\paragraph{Volume Extraction Methods}

Similar to the price extraction methods part, the price components from the raw LOB states shown in equation \ref{baselob} can be removed, which leaves the aggregated volume values at each non-empty bid and ask level. This would lead to the following extracted feature.

\begin{equation}
    \textbf{s}_t^{volume} := (v_t^{1,a}, v_t^{1,b}, ..., v_t^{10,a}, v_t^{10,b})^T \in \mathbb{R}^{20},
\end{equation}

As mentioned, there is room for improvement. If we sum up the quantities in the ask and bid side separately for the first ten non-empty price levels, we get the following.

\begin{equation}
    \textbf{s}_t^{\sum volume } := \left(\sum_{n=1}^{10}v_t^{n,a}, \, \sum_{n=1}^{10}v_t^{n,b}\right)^T \in \mathbb{R}^{2},
\end{equation}
This shows the number of shares in the observable region on the ask side and the bid side. If there are more shares on the ask side than on the bid side, this indicates that there are more buyers in the market, and so price is more likely to increase. Similarly, if there are more shares on the bid side than on the ask side, this indicates that there are more sellers in the market, and so price is more likely to decrease. To show this more clearly, one can simply subtract the two values and this is the calculation for \textit{volume delta} (VD).

\begin{equation}
    \textbf{s}_t^{VD} := \sum_{n=1}^{10}(v_t^{n,a} -v_t^{n,b})\in \mathbb{R},
    \label{vol}
\end{equation}

Again, a positive value indicates buying power and a negative value indicates selling power. \textit{Cumulative volume deltas} (CVD) take this idea further and show the difference in bid and ask volumes between consecutive timesteps. This reveals the change in equation \ref{vol} over time, which can identify a sudden increase in buying or selling power (very useful during important news releases) and can be calculated as the following.

\begin{equation}
    \textbf{s}_t^{CVD} := \left( \sum_{n=1}^{10}(v_i^{n,a} -v_i^{n,b}) - \sum_{n=1}^{10}(v_{i-1}^{n,a} - v_{i-1}^{n,b}) \, \Big| \, i \in [ t-9, t ] \right)^T \in \mathbb{R}^{10},
\end{equation}

There are many algorithms that use cumulative volume deltas; notable examples consist of volume delta reversal strategies \cite{kohout}, which identify a change of sign in the cumulative volume delta and trade expecting price to reverse direction.

\paragraph{Order Flow Extraction Methods}

\textit{Order flow} shows the real-time movement of orders entering the LOB. Unlike the mentioned extraction methods, this feature retains both price and volume data, which gives it the potential to provide more insight into supply and demand dynamics. The following explains how to calculate order flow from the LOB as described in \cite[Pages 6-7]{kolm}.\\

Order flow ($\textbf{OF}_t$) captures the change in the LOB state and, therefore, requires two consecutive tuples, i.e. at time $t$ and $t-1$. It is calculated in the following way:
\begin{equation}
    \text{aOF}_{t,i} := 
    \begin{cases}
        v_t^{i,a},  \,\,\,\,\,\,\,\,\,\,\,\,\,\,\,\,\,\,\,\,\, \text{if} \,\, a_t^i < a_{t-1}^i,\\
        v_t^{i,a} - v_{t-1}^{i,a}, \,\,\,\, \text{if} \,\, a_t^i = a_{t-1}^i,\\
        -v_t^{i,a}, \,\,\,\,\,\,\,\,\,\,\,\,\,\,\,\,\, \text{if} \,\, a_t^i > a_{t-1}^i,\\
    \end{cases}
\end{equation}
\begin{equation}
    \text{bOF}_{t,i} := 
    \begin{cases}
        v_t^{i,b},  \,\,\,\,\,\,\,\,\,\,\,\,\,\,\,\,\,\,\,\,\, \text{if} \,\, b_t^i > b_{t-1}^i,\\
        v_t^{i,b} - v_{t-1}^{i,b}, \,\,\,\, \text{if} \,\, b_t^i = b_{t-1}^i,\\
        -v_t^{i,b}, \,\,\,\,\,\,\,\,\,\,\,\,\,\,\,\,\, \text{if} \,\, b_t^i < b_{t-1}^i,\\
    \end{cases}
\end{equation}
where $\text{aOF}_{t,i}$ and $\text{bOF}_{t,i}$ are the ask and bid order flow elements for the $i$-th level (1 to 10 incl.) at time $t$. Order flow can be obtained through concatenation:
\begin{equation}
    \text{OF}_{t} := \begin{pmatrix} \text{bOF}_{t}\\ \text{aOF}_{t} \end{pmatrix} \in \mathbb{R}^{20},
\end{equation}
\textit{Order flow imbalance} takes this further as it reveals the disparity between the ask and bid order flow components. It can be derived as the following.

\begin{equation} \label{eq:ofi}
    \text{OFI}_{t} := \text{bOF}_{t} - \text{aOF}_{t} \in \mathbb{R}^{10},
\end{equation}

\subsubsection{Supervised Learning on Order Books}
This Section describes the recent work on the topic of forecasting small price changes (\textit{alpha}) using supervised learning on features extracted from the LOB. Tran et al. \cite[30.5, Pages 1407–1418]{tran}, Passalis et al. \cite[136, Pages 183-189]{passalis}, and M\"{a}kinen et al. \cite[19.12, Pages 2033-2050]{ymir} independently but commonly used classifiers to predict price movement, either labelling data into two classes (up or down) or three classes (up, down, or sideways). Labels were predominantly assigned by applying moving averages to price and then using thresholds. Although these methods remove noise from the data, they add modelling parameters that are not desirable in real-world trading applications. For example, the justifications for setting parameters to particular values are probably unreliable because these values are likely to change if there is more data. Kolm et al. \cite{kolm} address this by changing the problem to regression- a better choice as it removes unwanted assumptions but is more computationally demanding.\\

Kolm et al.'s \cite{kolm} design extracts alpha at multiple timesteps (\textit{horizons})- alpha is the expected change in price, usually after a very short period. Inspired by Cont et al. \cite[Pages 47-88]{cont} on stationary quantities derived from the LOB (order flow), networks trained on this outperformed networks that were trained from the raw LOB. Kolm et al. \cite{kolm} took this idea and found that using order flow imbalance as features was a further improvement in aiding the learning process. Regarding the data used for learning, it was standard to source high-quality LOB data from LOBSTER \cite{LOBSTER}.\\

\begin{figure}[h!]
\centering
\includegraphics[width = \hsize]{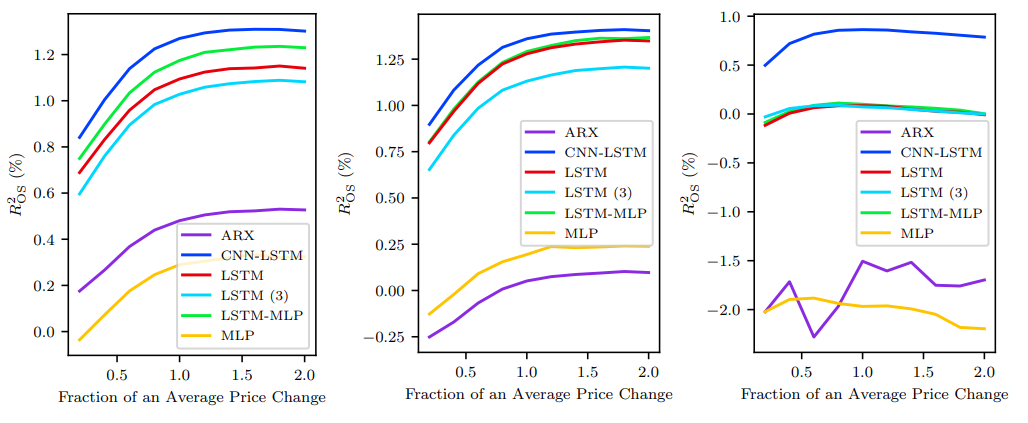}
\caption{Forecasting performance of different models using OFI (left panel), OF (middle panel), and raw LOB (right panel). Each panel shows the average out-of-sample $R^2$ against horizons represented as the fraction of an average price change per stock. These graphs can be found in \cite[Pages 18 and 35]{kolm}}
\label{fig:kolmresults}
\end{figure}

Figure \ref{fig:kolmresults} on the next page presents the results from \cite{kolm} of the forecasting performance of selective models against the ratio of horizon period over price change. Each model was trained while adopting a 3-week rolling-window \textit{out-of-sample} methodology across 48 weeks using a (1-week validation, 4-weeks training, 1-week out-of-sample testing) structure. The evaluation metric used is out-of-sample $R^2$ ($R^2_{\text{OS}})$. It is defined in \cite[Page 17]{kolm} as
\begin{equation}
    R^2_{\text{OS}, h} := 1 - \text{MSE}_{\text{m},h}/\text{MSE}_{\text{bmk},h},
    \label{eq:rsquared}
\end{equation}
for each \textit{horizon} $h$, where $\text{MSE}_{\text{m},h}\,\text{and}\, \text{MSE}_{\text{bmk},h}$ are the mean-squared errors of the model forecasts and benchmark, respectively. The benchmark used is the average out-of-sample return of the stock- 115 symbols were used, and the average across all are presented in figure \ref{fig:kolmresults}. Overall, the results show that models perform better when using OF and OFI as input rather than raw LOB and that CNN-LSTM extracts alpha most accurately, followed by LSTM-MLP, LSTM, and a stacked LSTM with three layers. MLP stands for Multi-Layer Perceptron, which is a stack of linear layers. The $R^2_{\text{OS}}$ (\%) is mostly positive, meaning that these networks beat the benchmark. 

\subsubsection{Reinforcement Learning on Order Books}
The related work done using reinforcement learning (RL) on the LOB to execute or simulate trades will be covered. The first large-scale experiments were conducted in 2006 by Nevmyvaka et al. \cite{nevmyvaka}, showing the potential of using RL on the LOB. As technology improved alongside increasingly available data and advanced algorithms, RL became more widely used in high-frequency trading. Spooner et al. in 2018 \cite{spooner} evaluated temporal-difference algorithms (including Q-Learning) in realistic simulations and observed promising performance. To address the trade-off between maximising profits and managing risk in trading, Mani et al. in 2019 \cite{mani} incorporated risk-sensitive RL. This agent's decisions are not only led by potential profits but are also influenced by potential loss, making it more robust to uncertainty. Despite this, it is considered risky to give an RL agent capital to work with as it lacks explainability, especially in a business context where explainability is essential. Vyetrenko et al. in 2019 \cite{Vyetrenko} address this for risk-sensitive RL strategies by representing the agent's decisions in compact decision trees.\\

Karpe et al. in 2020 \cite{karpe} use a Double Deep Q Learning network (DDQN) as well as other RL agents in a realistic LOB market environment simulation. It was astonishing to read that one of their RL agents, in certain scenarios, independently demonstrated the adoption of an existing method: the Time-Weighted Average Price (TWAP) strategy. In brief, the TWAP strategy, according to \cite{bybithelp}, aims to prevent sudden market volatility (large order impact) by dividing a large order into multiple smaller orders submitted at even time intervals to achieve an average execution price that is close to the actual price of the instrument.

\begin{figure}[h!]
\centering
\includegraphics[width = \hsize]{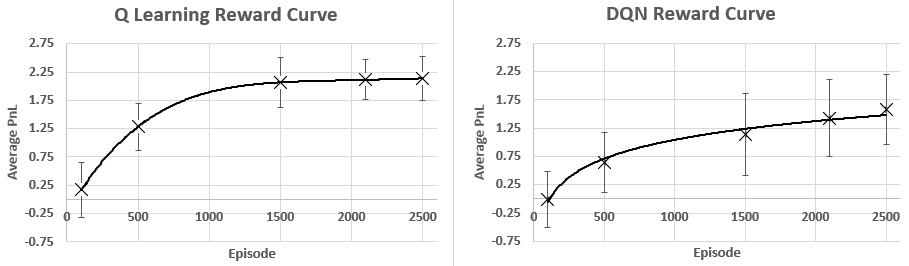}
\caption{Graphical performance comparison (Average PnL) of Q Learning and DQN inferred from a tabular version provided by Bertermann \cite[Page 29]{bertermann}; the error bars are standard deviations}
\label{fig:bertermannresults2}
\end{figure}

Bertermann in 2021 \cite{bertermann} implemented and compared many RL algorithms, including Q Learning and DQN, for high-frequency trading using the LOB. The features used are long, mid, and short-term mean-reverting signals. Figure \ref{fig:bertermannresults2} displays the results graphically inferred from the original tabular version \cite{bertermann} shown in figure \ref{fig:bertermannresults} in the Appendices section. From initial inspection, the Q Learning agent has better convergence, and the average profit accumulated after nearly 2,500 episodes of training the Q Learning agent made 35\% more than a DQN agent under the same conditions. On top of the standard deviation being 36.3\% smaller near 2500 episodes, it implies higher profitable consistency. It is clear that Q Learning is the better algorithm for trading according to these results, which further supports that DQN suffers from instability and overestimation, as mentioned in the Background Section. However, it is important to note that DQN has more changeable parameters, and it could be the case that the parameters were sub-optimal in this study.

\subsubsection{Evaluating Trading Agents}
There are other ways to evaluate trading agents beyond simply looking at how much money they make during testing (\textit{PnL}). Here is a brief list of common performance metrics accumulated from \cite{singh, auquan}.

\begin{itemize}
    \item Daily Average Profit/Return: the average of profit or return at the daily level
    \item Daily Average Volatility: the standard deviation of return at the daily level
    \item Average Profit/Loss: the ratio of average profit over average loss
    \item Profitability (\%): the percentage of trades that result in a profit
    \item Maximum Drawdown: the largest peak to trough in the equity curve

\end{itemize}
The Sharpe and Sortino ratios will be omitted as they are not insightful for comparing different agents over the \textbf{same} test data. They assess the average excess return beyond a benchmark over volatility (downside volatility for Sortino ratio) and, therefore, are proportional to the return while everything else is constant.

\section{Design, Implementation, and Testing}

This Section covers the design decisions and the implementation of the solution, as well as mentions the process of testing the models. The link to the code repository is \textit{https://github.com/KotiJaddu/Masters-Project}.\\

The goal is to investigate the combination of deep learning on the order books and reinforcement learning for profitable trading. Two successful studies were selected (one supervised learning and one reinforcement learning) that needed each other in order to be complete, although there was no hesitation in bringing inspiration from other sources. From the different ideas explored in the related literature section, the approach taken forward was adopting Kolm et al.'s \cite{kolm} very promising alpha extraction to cover the "deep learning on the order books" and sending those outputs into some of the RL agents evaluated by Bertermann \cite{bertermann} to cover the "reinforcement learning for profitable trading". Kolm et al.'s alpha extraction needs Bertermann's reinforcement learning to productise their work, and likewise, Bertermann's reinforcement learning requires quality features compared to lagging mean-reverting signals (despite observing remarkable results).

\subsection{Data Collection}
This Section discusses the design of the data collection pipeline and evaluates the quality of the collected data. The models to be trained can only be as good as the data fed into them, so a reasonable amount of time was invested into this.\\

The widely used dataset in relevant literature is sourced from LOBSTER \cite{LOBSTER}. Although this would also fit well into this work, I wish to integrate the models into my personal retail trading platform (CTrader FXPro \cite{fxpro}) to execute automated trades for realistic simulated testing. This is an award-winning broker with a coding interface that allows access to their LOB. As price action in CTrader is reflected by the orders of its users, the networks will be trained on the LOB data from the platform instead of sourcing it from LOBSTER to maximise accuracy.\\

\begin{figure}[h!]
\centering
\includegraphics[width = \hsize]{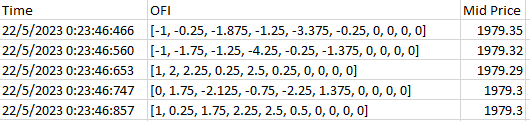}
\caption{First five records of the CSV file containing (Time, Order Flow Imbalance, Mid Price) data}
\label{fig:csv}
\end{figure}

CTrader FXPro gives users access to the current state of the LOB but does not provide historical data. Hence, LOB data was collected as early as possible (23/5/2023), saving each LOB state as well as the mid-price per timestep to a CSV file for easy access. Timesteps are defined as changes to the observable LOB states. Storing raw LOB data could bring ethical issues, so the code on CTrader extracts the order flow imbalance features, which is what will be used as input to the networks, and stores that into the CSV file instead as in figure \ref{fig:csv}. The first ten OFI levels were used by Kolm et al. \cite{kolm} and should be sufficient to carry out this work.

As this data extraction should run continuously for ten weeks, this was set up on a Virtual Private Server (VPS) hosted by TradingFX VPS \cite{fxvps}, and the files were transferred to local storage every weekend to mitigate the loss upon any technical issue. With regard to the financial instruments that will be scraped, five popular assets were chosen: DE40, FTSE100, EUR/USD, GBP/USD, and XAU/USD (Gold). This covers indices, foreign exchange pairs, and metal. 
\subsubsection{Analysing Collected Data}

Now, analysis will be conducted on the data collected to evaluate its quality. This includes presenting the basic attributes of the data such as mean, standard deviation, and the number of data points. A histogram will also be plotted to look for normality by fitting a bell curve for each dimension in the order flow imbalance vector for each instrument. Normality infers a symmetrical dataset (equal spread of data points on either side of the mean), which improves convergence during learning.\\

All data was collected on every working day from 23/5/2023 to 3/7/2023 (6 weeks) and from 25/7/2023 to 21/8/2023 (four weeks). There was a technical issue which caused missing data from 4/7/2023 to 24/7/2023, but ten weeks of data was collected in total nevertheless. There is an exception with FTSE100, which is missing one day due to the UK spring holiday (29th of May). 

\begin{table}[htbp]
  \centering
  \caption{Basic Attributes of the Collected OFI Data}
    \begin{tabular}{lccccc}
          Attribute  & XAUUSD & GBPUSD & EURUSD & FTSE100 & DE40 \\
        \hline
          Total Data Points & 13,356,795& 6,177,143&4,513,533&  1,647,380 & 1,535,992\\
          Daily Avg. Count  & 267,136& 123,543& 90,271&  33,620 &30,720 \\
          Mean OFI value& 0.1218& 0.0846 & 0.0775& 0.1052 &0.0203\\
          \hline
          Standard dev. & 0.4197& 0.4060& 0.4136& 0.4848 &0.3727\\
          \% Positive &48.0 & 47.3 &48.0  &37.6& 41.3\\
          \% Negative &20.4& 26.0 & 28.3 &23.5& 32.1\\
          \% Zero&  31.6 &26.7 &23.7 &38.9 &26.6\\
        \hline
    \end{tabular}
  \label{tab:ofiattributes}
\end{table}

Table \ref{tab:ofiattributes} shows the basic attributes of the collected data. The number of data points will be sufficient to train the small-scale models for each instrument. It is interesting to see that all the means are positive, indicating that the ten weeks of data recorded could be bullish-biased. This is supported by all the OFI containing almost twice as many positive values as negative values, although this might not necessarily directly correlate to positive price movement. However, the standard deviations being almost five times more than the means infers that there is sufficient coverage of negative values in the data.\\

To better understand the spread of the data, figures \ref{fig:ofidist_xauusd}-\ref{fig:ofidist_ftse100} in the Appendices Section show the histograms of OFI values (normalised and then scaled) at each of the ten levels for each instrument as well as a fitted normal distribution on top. Overall, the spread is favoured towards the right side of the mean, and the fitted normal distribution is somewhat appropriate as one could observe a peak near the mean and then tails towards the extremities. These graphs show sufficient coverage of positive and negative values, enough to train the network for alpha extraction. The OFI distributions for DE40 (figure \ref{fig:ofidist_de40}) are the best as almost all levels show a valid fit with the exception of levels 1 (bimodal), 5 (skewed), and 6 (bimodal).\\

Tables \ref{tab:ofimean}-\ref{tab:ofizer} in the Appendices Section go into more detail showing the mean, standard deviation, and percentage of positive, negative, and zero values for each OFI level in the collected data for every instrument. As expected, the proportion of zero values increases the higher the OFI level because more updates to the LOB happen at lower levels near the mid-price, especially at levels 2 to 6.\\

Both supervised learning and reinforcement learning components will use 80\% of the data for training, 10\% of the data for validation (for hyperparameter tuning), and the final 10\% of the data for testing.

\subsection{Supervised Learning}
This Section discusses the decisions made during the integration of Kolm et al's \cite{kolm} alpha extraction into this work. Overall, order flow imbalance features will be collected, which will be used as input to a supervised learning regression model that will predict the change in mid-price (alpha) for the next six horizons. The reason for choosing six is that Kolm et al. \cite{kolm} observed that price prediction accuracy falls off after six horizons. The output of the code is to save the model to disk for use in the reinforcement learning part. At the end of this Section, the quality of alphas to be set as labels to OFI features are evaluated to ensure good coverage.\\

The learning of a model for each instrument will be written as a reusable function using Python- the Python programming language was chosen because of the vast number of machine learning libraries and available documentation. Furthermore, the PyTorch library \cite{pytorch} will be used, which has more functionality than needed for this work. With regards to the code, it will be well commented to ensure readability such that any developer can make further modifications with ease. It will also make use of the GPU with CUDA \cite{cuda} during training to save time.

\subsubsection{Pipeline}

\begin{algorithm}
\caption{Supervised Learning Algorithm}\label{alg:supervisedlearning}
Apply preprocessing to data\\
Split data ratio (7: 1: 2) into training $T$, validation $V$, and testing $H$ sets\\
Initialise hyperparameters\\
Initialise model $M$ and optimiser $\textit{Opt(Method, Learning Rate)}$\\
$k \gets 0$ \, \# for early stopping\\
\Repeat{$k =$  maximum patience tolerable or maximum epochs reached}{

    \Repeat{$T$ fully iterated}{
        $x, y \gets \textit{next batch from} \,T$ \, \# features $x$ and labels $y$\\
        $y' \gets M(x)$ \, \# \textit{predictions}\\
        $\textit{Loss} \gets L(y', y)$\\
        $\textit{Opt.backpropagate(Loss)}$ \, \# updates weights and bias values in $M$\\
    }
    \If{error of $M$ on $B$ has improved from previous best} {
        $k \gets 0$\\
    }\Else{
        $k \gets k + 1$\\
    }
   
}
Evaluate $M$ with $H$ and process results
\end{algorithm}

Algorithm \ref{alg:supervisedlearning} will be the pipeline as explained earlier with a further enhancement from Kolm et al. \cite{kolm}. It is normally a difficult task to figure out how many epochs are needed to train the model for optimal results. If the number of epochs is too small, then the model will be underfitting, and if the number of epochs is too large, then the model will be overfitting- both of which are undesirable, and a lot of time-consuming trialling is required. Early stopping will terminate the training when it thinks the model is starting to overfit. It does this with a validation set, as in algorithm \ref{alg:supervisedlearning}, by checking how many epochs in a row the model has not improved since its previous best score, and if this number of epochs exceeds the maximum patience tolerable, then the learning will terminate. However, a problem arises when the maximum number of epochs is not limited, which is very long training times. A solution is to set a limit to the maximum number of epochs tolerable.\\

As recommended by Kolm et al. \cite{kolm}, L2 regularisation was adopted to prevent overfitting. This adds a small penalty term to the loss function and encourages the weights to be small. From initial experimentation, using LSTM and LSTM-CNN layers was not beneficial compared to using just a multi-layer perceptron (MLP). The reason for this could be that these complex networks are difficult to train with limited hardware. Although not as successful as LSTM and LSTM-CNN networks, Kolm et al. \cite{kolm} show that a MLP also produces promising results, so this will be the backbone architecture of the regression networks.\\

There are many hyperparameters in algorithm \ref{alg:supervisedlearning} that require tuning. These are batch size, optimiser method, learning rate, network architecture, loss function, maximum epoch number, and early stopping patience parameter. Although optimisation is an important topic for maximising model accuracy, the methodology for tuning these variables will be covered in the next Section - Optimisation.\\

Regarding the preprocessing on line 1, limiting the data to remove periods of low trading participation (pre and post-market) came to mind. However, the OFI should also indicate low volatility, which the network should pick up. Keeping all trading periods in the data is also a suggestion in the further works of Karpe et al. [26] for their work, so nothing was removed.

\subsubsection{Analysing Alphas from Collected Data}

The preprocessing also involves calculating alphas from the data and setting them as labels to the OFI features. Please see figure \ref{fig:alphas} in the Appendices Section for tabular outputs. The attributes of the alphas will be analysed to evaluate the spread of the labels in the data. The process will be similar to the previous evaluation of data and will be applied to each instrument.

\begin{table}[htbp]
  \centering
  \caption{Basic Attributes of the Calculated Alphas}
    \begin{tabular}{lccccc}
          Attribute  & XAUUSD & GBPUSD & EURUSD & FTSE100 & DE40 \\
         \hline
          Mean value (Price) & -2.85e-05&  1.15e-08 &-4.25e-09 &-1.42e-03 &-1.79e-03\\
          Standard dev. (Price) & 4.80e-02 & 3.16e-05 & 2.74e-05 & 5.30e-01  &1.40e+00\\
          Tick Size & 0.01 & 0.0001 & 0.0001 & 0.1 & 0.1\\
          Mean value (Pips) & -2.85e-03&  1.15e-04 &-4.25e-05 &-1.42e-02 &-1.79e-02\\
          \hline
          Standard dev. (Pips)& 4.80e+00  & 3.16e-01 & 2.74e-01 & 5.30e+00  &1.40e+01\\
          \% Positive &37.3 & 38.7&  37.2  &39.5& 35.2\\
          \% Negative &37.9 & 38.6&  37.4 & 39.5& 35.2\\
          \% Zero&  24.8&  22.7 & 25.4 &21.0 &29.6\\
        \hline
    \end{tabular}
  \label{tab:alphaattributes}
\end{table}

Table \ref{tab:alphaattributes} presents the attributes of the calculated alphas. The tick size is the smallest movement that the best bid or ask level can move in price units, which is equal to 1 pip. Converting price movements to pips will allow to standardise the values for comparison across all instruments. From this data, the instruments can be ordered (descending) according to standard deviation, which is the volatility and risk: DE40, FTSE100, XAUUSD, GBPUSD, and EURUSD.\\

Furthermore, the table shows that the mean alpha movement in terms of pips is close to 0, and on top of a much larger standard deviation, there seems to be good coverage of positive and negative values. This is further supported by the proportion of positive and negative values being almost identical for every instrument, which is ideal because the network can learn bullish and bearish behaviour equally. It is also important for the network to learn when price will not move, so a good proportion of zero movement is also very useful. Tables \ref{tab:alphamean}-\ref{tab:alphazer} in the Appendices Section go into more detail for each horizon level in the collected data for every instrument. An interesting pattern followed by all instruments except XAUUSD is that price usually oscillates back and forth between adjacent price levels. This can be seen through the percentage of zero alphas peaking at every other horizon in table \ref{tab:alphazer}.

\subsection{Reinforcement Learning}
This Section discusses the decisions made during the integration of Bertermann's \cite{bertermann} RL agents (Q Learning and Deep Q Learning) into this work. A Double Deep Q Learning Network (DDQN) will also be investigated as it showed successful results according to Karpe et al. \cite{karpe} and addresses the problems with the DQN as discussed in Section (1.2.4).\\

The code will be written in the Python programming language using PyTorch wherever applicable, as mentioned in the previous Section. There will be three scripts dedicated to representing each agent: Q Learning, DQN, and DDQN. This separates the agents such that any can be used for future work, which is difficult with the alternative approach consisting of merging all agents into one condensed file and passing the agent to use through the command line. However, this alternative approach would be more efficient as each file includes the learning process and testing method on unseen data to evaluate the performance. Algorithms \ref{alg:qLearning}, \ref{alg:dqn}, and \ref{alg:ddqn} will be used to implement the RL agents. There are many hyperparameters that will need to be tuned, which will be discussed in the next Section.\\

There are two methods for using market data to train these RL agents: using offline data (see \textit{backtesting} Section 2.4.1) and using online data (see \textit{forward testing}). Offline data involves iterating through historical data for training, which is faster to process compared to using online data that uses real-time data given from an environment. As the supervised learning component requires the storing of historical data, offline data is therefore available to simulate the markets for learning. The implementation involves coding the backtesting process, using algorithm \ref{alg:backtesting}, in a separate Python script, which will be used by all RL agents. Taking inspiration from OpenAI's Gym Environment \cite{gym}, the backtesting is formatted as an environment (in the utils.py script) that sends the agent the next state and reward when given the current state and action while applying retail-level commissions to the profit or loss for each trade. Forward testing is important for realistic simulation, so CTrader's built-in forward-tester (algorithm \ref{alg:forwardtesting}) will be used. This involves creating a web application using Flask \cite{flask}, which will use the models to send signals to the platform upon receiving POST requests. Another way was to edit and read text files, but this would slow the process as both applications cannot access the same file at once.\\

There are two ways to use the historical data for training. The first is using the predicted alphas from the supervised learning component (alpha extraction) as input to the RL agent. The second is calculating the actual alphas stored in the historical data and using that as input. The first method allows the RL agent to train to the inaccuracies of the alpha extraction model, whereas the second trains independently to the alpha extraction- which relies on the alpha extraction to be accurate for the RL agent to fully utilise its training. The second method was selected to avoid training bias. Due to limited data, there will be an overlap of data used to train both the alpha extraction and RL agents. This means that the alpha extraction will be more accurate when passing through data it was trained on, and its accuracy will likely differ on unseen data- this will confuse the RL agents. Furthermore, keeping the training process independent allows for the alpha extraction to be changed or improved without the RL agents being affected.\\

The agent will always hold an active position in the market at all times whether, it be a buy or a sell position. This is to keep the agent simple and show potential for this approach. The agent can only alternate the direction upon receiving a reversal signal; therefore, it should have enough knowledge to be able to weigh the potential reward for reversing the current position, given the transaction costs in doing so. Hence, the state space should also contain the state of the current position on top of the alphas as input so it can figure out whether it is worth reversing the current position. There will be seven states in total (six alphas and one current position). This is an extension to Bertermann's \cite{bertermann} setup.\\

From initial experimentation, using exploration decay was found to be important, which decreases the amount of exploration done by the agent over time while emphasising high exploration during early stages of training. This was recommended by Bertermann \cite{bertermann}. On the other hand, a decay in learning rate did not show any improvements in learning and, in most cases, worsened performance. L2 regularisation was also incorporated in the DQN and DDQN models to prevent overfitting by encouraging the weights to be small.

\subsection{Testing}
This Section covers the methodology of testing each model to evaluate its performance fairly for comparison. Backtesting and forward testing will be explained.
\subsubsection{Backtesting}

Trading strategies or automated trading bots need to be evaluated in a risk-free but realistic environment before it is validated for live trading with real money. \textit{Backtesting} is the process of simulating the market with historical data to track the results of buy/sell signals according to a strategy. The actual outcomes had these signals been taken as trades will be used to assess the agent's performance at the end of the simulation.\\

Algorithm \ref{alg:backtesting} shows a backtesting routine for our agent that alternates between continuously holding an active buy or sell position on an instrument. All the trade signals, as well as their results, are logged in $T$, which will be used to calculate the performance metrics needed to evaluate the agent. As CTrader does not provide historical LOB data, backtesting will need to be implemented independently using code. In line 14, transaction costs include commissions and spread (the difference between the best ask price and best bid price).

\begin{algorithm}
\caption{Single Position HFT Backtesting Algorithm}\label{alg:backtesting}
Load agent $A$\\
Load backtesting data $D$ as a list of tuples with form $($\textit{OFI}, \textit{Mid-price}$)$\\
Initialise empty trade log $T$\\
Initialise previous action $a' \gets \textit{NULL}$\\
Initialise current trade $t \gets \textit{NULL}$\\
Initialise current state $s \gets \textit{next}(D)$\\
\Repeat{$D$ is fully iterated}{
    $a \gets A(s) $ \, \# \textit{buy or sell signal} \\
    \If {$a = a'$}{
        Update $t$ using $s$\\
    }\Else{
        Append non-NULL $t$ to $T$\\
        Reinitialise $t$ with $(s, a)$ and transaction costs\\
    }
    $s \gets \textit{next}(D)$ \\
    $a' \gets a$
}
Use $T$ to calculate performance metrics
\end{algorithm}

There are many events that are difficult to replicate in this environment due to shifts in market volatility from news releases and events. Some are the following:
\begin{itemize}
    \item Fluctuating spreads from traders being either passive or active during news
    \item Slippage is when the execution price is not the expected price of entry
    \item Being partially filled or not being filled if the entry price is hit but reverses
\end{itemize}
\subsubsection{Forward Testing}

When a trading strategy has been proven to be promising, usually from good performance during backtesting, one will consider passing their agent through \textit{forward testing}. This is a form of testing that connects an agent to a platform, like CTrader, where it can process live unseen data and submit real orders. It can trade with either real money or fake money (known as \textit{paper trading}).\\

This is more realistic as the platform will notify the agent when an order has been successfully filled (with the exception of paper trading as there is no real order), filled at a different price compared to the expected price of entry (slippage), and affected from fluctuating spreads. The results from forward testing will, therefore, be more accurate compared to backtesting. From inspection, if forward testing is more accurate compared to backtesting, then why do backtesting at all? Why not only do forward testing? It is because forward testing is more time-consuming as it is using live data compared to iterating through historical data like in backtesting.\\

Similar to before, Algorithm \ref{alg:forwardtesting} shows a forward testing method for our agent that alternates between continuously holding an active buy or sell position.

\begin{algorithm}
\caption{Single Position HFT Forward Testing Algorithm}\label{alg:forwardtesting}
Load agent $A$\\
Acquire live data feed $s$ as a tuple $($\textit{OFI}, \textit{Mid-price}$)$\\
Initialise empty trade log $T$\\
Initialise previous action $a' \gets \textit{NULL}$\\
\Repeat{terminated}{
    $a \gets A(s) $ \, \# \textit{buy or sell signal} \\
    \If {$a$ is not $a'$}{
        Append result of $t$ if it exists to $T$\\
        Close position $t$ if it exists and open position $t$ aligned with $a$
    }
    Wait for an update from the live data feed and store the new state in $s$ \\
    $a' \gets a$
}
Use $T$ to calculate performance metrics
\end{algorithm}
The main changes consist of using the live data feed instead of historical data (line 2), closing and opening a new trade on the trading platform (line 9), and waiting for a new incoming update from the data feed (line 11).

\section{Optimisation}
This Section discusses any experimentation and optimisation undertaken to improve the solutions for the supervised learning and reinforcement learning components. This will primarily be via hyperparameter tuning to improve the models after starting with settings proposed by Kolm et al. \cite{kolm} and Bertermann \cite{bertermann} as a baseline for certain parameters. Hyperparameters already discussed in the Design Section will reappear for completion. The outcome of the tuning will be presented in the next Section.

\subsection{Supervised Learning Tuning}
Here is a list of all the hyperparameters that will need to be tuned for the alpha extraction component.
\begin{itemize}
    \item Network architecture $\rightarrow$ includes the type of layers that should be used in the network, the number of hidden layers, the number of nodes in each hidden layer, and the activation function to apply after each layer. These control how data is processed, which needs to be appropriate to the task. They also configure the capacity of the network- too large capacity leads to overfitting, and too small capacity leads to underfitting.
    \item Optimiser method $\rightarrow$ defines the way to optimise the model's parameters during training to reduce the error and converge towards the optimal configuration. Choosing the right optimiser is important as it impacts the convergence speed and the final performance of the model.
    \item Learning rate $\rightarrow$ is the step size determining how much the model's parameters are adjusted- too small rates result in slow convergence, whereas too large rates will fail to converge due to overshooting the optimal solution.
    \item Batch size $\rightarrow$ is the number of records in the training data used to update the model's parameters in one go. This adds noise to the model, which helps it generalise to unseen data. Small batch sizes inject too much noise into the model, which slows training, while large batch sizes reduce the regularisation effect and are computationally expensive.
    \item Loss function $\rightarrow$ is used to compare the predicted value to the actual value and compute an error. An inappropriate function will result in the model learning and prioritising incorrectly.
    \item Early Stopping Patience $\rightarrow$ is used to stop the training when the model begins to overfit. If this value is too low, then the model can underfit, but if it is too large, then the model can overfit the data.
    \item Maximum number of Epochs $\rightarrow$ the number of epochs before automatically terminating the training process. This is to avoid very long training times when only relying on early stopping.
\end{itemize}

\subsubsection{Justifying Values for Certain Parameters}
There are certain parameters where their optimal values can be justified using existing knowledge. Hence, they will not need to be tuned unless they are a starting point from existing literature. Here is a list of hyperparameters where their values are justified among popular alternatives.

\begin{itemize}
    \item Type of layers (Linear, LSTM, LSTM-CNN) $\rightarrow$ Initial experiments showed that these layers added no value on top of linear layers- perhaps due to limited hardware. Kolm et al. \cite{kolm} show that linear layers also give promising results.
    \item Hidden Layer Count and Nodes ([3, 4, 5], [1024, 2048, 4096]) $\rightarrow$ Kolm et al. \cite{kolm} use four hidden layers, which should be sufficient, and the number of nodes will need to be found through exploration via tuning.
    \item Learning Rate (0.00001, 0.0001) $\rightarrow$ Kolm et al. \cite{kolm} use 0.00001 as their learning rate, so this is used as a baseline.
    \item Activation function (ReLU, Tanh, Sigmoid) $\rightarrow$ There is a saturation problem with Tanh and Sigmoid: as the inputs to the functions tend towards positive infinity or negative infinity, the gradient approaches zero, leading to vanishing gradients. ReLU does not have this because it is a linear function when the input is positive. It is also easier to compute ReLU, making it computationally efficient.
    \item Batch size (128, 256, 512) $\rightarrow$ Kolm et al. \cite{kolm} use 256 as their batch size, so this will be the starting point.
    \item Loss function (MSE, RMSE, MAE) $\rightarrow$ RMSE (Root mean squared error) is used to make the error more interpretable, but this is not important for training the model. MAE (Mean absolute error) is less sensitive to outliers, but our data is observed from a real data feed, so there are no real outliers. If it is possible to observe a data point that is outside of the expected distribution, then the model should be aware of this. It is common to use MSE (Mean squared error) for training regression models.
    \item Optimiser method (SGD, ADAM) $\rightarrow$ ADAM (Adaptive moment estimation) is an extended version of SGD (Stochastic gradient descent) by adapting the learning rate for each parameter based on the gradients from the previous optimisation step. This means that ADAM is less sensitive to the initial learning rate choice as it will continuously change throughout training to a more appropriate value. ADAM also handles different gradient scales better than SGD because it uses the first and second moments of the gradients. Kolm et al. \cite{kolm} also use ADAM.
    \item Early Stopping Patience (5, 10) $\rightarrow$ 5 is recommended by Kolm et al. \cite{kolm} so this will be the starting point. 
    \item Maximum number of Epochs (50, 100, 150) $\rightarrow$ 100 is chosen from tolerance.
\end{itemize}

\subsubsection{Methodology for Tuning Parameters}
There are many ways to optimise the hyperparameters, but since a reasonable amount of time can be dedicated to tuning, the simplest yet robust method will be implemented: grid search. This involves exhaustively iterating through all defined configurations of hyperparameters and using the combination that has the best validation score. Table \ref{tab:supervisedlearning_tuning} contains the proposed configurations for each hyperparameter, and this will be executed for all five instruments. The hyperparameters that will be tuned are the parameters with limited knowledge of their optimal values and, therefore, require experimentation.

\begin{table}[htbp]
  \centering
  \caption{Supervised Learning Hyperparameter Configurations for Grid Search}
    \begin{tabular}{lccccc}
          Hyperparameter               & Configurations                            \\
        \hline
          Hidden Layer Count and Nodes               & [512] * 4, [1024] * 4, [2048] * 4  \\
          Learning Rate   & 0.00001, 0.0001                   \\
          Early Stopping Parameter   & 5, 10                  \\
          Batch Size   & 128, 256, 512                   \\
        \hline
    \end{tabular}
  \label{tab:supervisedlearning_tuning}
\end{table}

There are a total of 36 combinations, and each run takes between 10 and 30 minutes. Taking 20 minutes as an average per run and repeating this for four other instruments leads to a total of 60 hours (20*36*5/60).

\subsection{Reinforcement Learning Tuning}
Here is a list of all the hyperparameters that will need to be tuned for the reinforcement learning component. As there are three agents, beside each parameter in the list below is/are the agent(s) that it belongs to. Please note that DQN and DDQN will be tuned together. There will be repeated hyperparameters from the supervised learning tuning because there are also neural networks in this component. For such parameters, please refer to the previous Section because the same strategies have also been adopted in this part unless mentioned otherwise.

\begin{itemize}
    \item (All) Maximum and minimum exploration rate $\rightarrow$ is the probability of making a random action during learning and aims to explore other options, rather than using the learned model, to find better rewards. The exploration rate should be the highest towards the beginning of the learning process and should gradually decrease towards a minimum. If the exploration rate is too low, then the agent will not have a chance to explore the environment enough. If the exploration rate is too high, then the convergence will be slower due to unstable learning, as the agent will be taking too many random/sub-optimal actions.
    \item (All) Exploration rate decay $\rightarrow$ is the rate at which the exploration rate decays throughout training (exponentially). If the decay decreases the exploration rate too quickly, then the agent will not have enough exposure to learn from alternative decisions. If the decay decreases the exploration too slowly, then the agent may take longer to converge due to unstable learning, as mentioned. The exploration rate at each episode is calculated as ($\textit{rate\_decay} ** episode$) and then set to the minimum exploration rate if exceeded.
    \item (All) Discounted future reward factor $\rightarrow$ is the importance given to future rewards during learning. A value closer to 0 makes the agent focus on immediate rewards, whereas a value closer to 1 causes the agent to focus on long-term rewards, which encourages more strategic decision-making.
    \item (All) Number of episodes $\rightarrow$ is the number of times the training set has been iterated. Each iteration through the training set is not the same because the agent will most likely take different actions leading to different rewards and, therefore, learn something new. If this number is too low, then the agent will not learn enough from the environment. Conversely, if this number is too big, then it is very time-consuming and has diminishing returns.
    \item (All) Reward function $\rightarrow$ is the function that returns a value (reward) when the agent is in a specific state. It is used to teach the agent whenever it does something correctly to get into a desirable state or incorrect to get into an undesirable state and hints the agent on what to prioritise as well as what to ignore.
    \item (All) Learning rate.
    \item (Q) State space coverage $\rightarrow$ is the state space covered by the Q table. If the state space coverage is too small, then it is likely to encounter states that are outside this region, which will need to be ignored, and this is not ideal if this occurs often. If the state space covered by the Q table is too large, then there will be redundant space in the Q table that will never be filled- ultimately wasting computational resources.
    \item (Q) Number of buckets $\rightarrow$ is the precision of separating the state space into buckets. If this value is too low, then it is likely that distant regions in the state space that should be separated will be merged into the same bucket. If this value is too high, then it is computationally expensive, and it will require a lot of time to accurately fill out all the values in the table.
    \item (DQN, DDQN) Batch size.
    \item (DQN, DDQN) Experience replay buffer size $\rightarrow$ is the size of the experience replay buffer. If this is too small, then it can cause overfitting to recent data as it can store a limited number of experiences. If this is too large, then it can be computationally expensive to store the data. Furthermore, the buffer is more likely to contain experiences that are outdated because the agent has improved a lot since those experiences were added to the buffer, which leads to redundant learning.
    \item (DQN, DDQN) Network architecture.
     \item (DQN, DDQN) Optimiser.
    \item (DQN, DDQN) Target update frequency $\rightarrow$ corresponds to the number of steps each time the target network updates its parameters to share the parameters of the policy network. If the update frequency is too low, then the target Q values for updating the policy network become outdated, leading to slow convergence. If the update frequency is too high, then the learning will become unstable because the policy network will try to catch up to a constantly changing target network.
    \item (DDQN) Target update weight $\rightarrow$ is the rate at which the target network's weights and biases are updated using the policy network's parameters. If this update weight is too small, then the target network will lag very behind compared to the policy network, causing slow convergence. If the update weight is too large, then the target network is more influenced by the policy network, which can cause learning to become unstable. The policy network will try catching up to a constantly changing target network.

\end{itemize}

\subsubsection{Justifying Values for Certain Parameters}

There are certain parameters where their optimal values can be justified using existing knowledge. Hence, they will not need to be tuned unless they are a starting point from existing literature. Here is a list of hyperparameters where their values are justified among popular alternatives.

\begin{itemize}
    \item (All) Maximum exploration rate (0.8, 0.9, 1.0) $\rightarrow$ As the initialised model before learning does not contain any relevant information regarding the task, the probability for the agent to execute random actions to learn in the environment should be the maximum possible value, which is 1.
    \item (All) Minimum exploration rate (0.0, 0.1, 0.2) $\rightarrow$ After the model has been exposed to the environment for enough episodes, the model should still make random actions to continue learning, but it should be low enough to stabilise learning. A 10\% chance for a random action to occur is reasonable when the model is beginning to stabilise.
    \item (All) Exploration rate decay $\rightarrow$ The minimum exploration rate should be reached after 70 out of 80 episodes so that it can use the remaining ten episodes to consolidate. Please see the number of episodes parameter below. According to the rate decay equation in the previous subsection, this means that the exploration decay constant must be 0.935 (3 sf).
    
    \item (All) Discounted future reward factor (0.9, 0.95, 0.99) $\rightarrow$ As profits are accumulated in the near future but not too far into the future, 0.99 might take the agent a very long time to converge. After 100 updates to price, the reward will still be contributing roughly 37\% to the state action Q value ($0.99^{100} \approx 0.37$). Bertermann \cite{bertermann} uses 0.9, but this will be further explored through tuning.
    \item (All) Number of episodes (40, 80, 120) $\rightarrow$ This is chosen from the amount of time waiting for the learning process to finish. This equates to two iterations of the entire training dataset (80 episodes), which takes roughly 45 minutes.
    \item (All) Reward function (PNL change, Account balance, \% profitability) $\rightarrow$ Each trade should be treated and taken independently. PNL (profit and loss) change is suitable because it returns the reward gained from the previous timestep, which leads to faster convergence. Account balance and \% profitability cause dependency issues as they can increase or decrease globally throughout the training process, leading to delayed learning.
    \item (All) Learning rate (0.1, 0.01, 0.001) $\rightarrow$ Bertermann \cite{bertermann} uses 0.001, so this will be a starting point.
    \item (Q) State range covered by Q Table (cover all, cover more, cover less) $\rightarrow$ If the Q table contains all the state space from the training data, then there will be many cells in the Q table that are empty because of outliers (0.5\% of the data). This will waste computational space. The 0.5\% of data towards the extremities was removed, which is roughly 2.5 standard deviations from the mean. It is better for an agent to say it does not know than to give an estimate for data points it has not been trained on.
    \item (Q) Number of buckets per state (3, 5, 7, 9) $\rightarrow$ This separates each alpha into one of five categories: large bearish bias, small bearish bias, no bias, small bullish bias, and large bullish bias. This should be sufficient information for the Q learning agent to make good decisions. As there are six horizons in the input data, two states for the current trade state, and two possible current actions, the Q table will have $5^{6} * 2^{2} = 62500$ cells.
    \item (DQN, DDQN) Batch size (64, 128, 256) $\rightarrow$ Bertermann \cite{bertermann} did not use experience replay, so there is no starting point. This will have to be trialled through tuning.
    \item (DQN, DDQN) Experience replay buffer size (4000, 6000, 8000, 10000) $\rightarrow$ A record of experience should stay in the buffer for 10,000 steps before it is claimed outdated. This is because an agent should make improvements every 10,000 steps, given that there are about 30,000 to 100,000 steps per day in the data.
    \item (DQN, DDQN) Hidden Layer Count and Nodes ([1, 2, 3], [16, 32, 64]) $\rightarrow$ The configurations for the hidden layer count and nodes are smaller compared to alpha extraction because classification with two outputs is simpler than regression with six outputs. This will need to be found through tuning.
     \item (DQN, DDQN) Type of Layers (Linear, LSTM, LSTM-CNN) $\rightarrow$ The inputs to these networks are much simpler compared to alpha extraction, so the networks will only need to use linear layers.
    \item (DQN, DDQN) Activation Function (ReLU, Tanh, Sigmoid) $\rightarrow$ The output layer in these networks has two nodes corresponding to probabilities of buying and selling. As these are probabilities, they need to be between 0 and 1. Sigmoid is a function that returns a value between 0 and 1, so it is suitable as the activation function for the output layer.
    \item (DQN, DDQN) Optimiser (SGD, ADAM).
    \item (DQN, DDQN) Target update frequency  (1, 5, 10) $\rightarrow$ Will have to be found through tuning.
    \item (DDQN) Target update weight (0.1, 0.2, 0.3) $\rightarrow$ Will have to be found through tuning.
\end{itemize}

\subsubsection{Methodology for Tuning Parameters}
Grid search was used to tune the remainder of the hyperparameters. Table \ref{tab:reinforcementlearning_tuning} contains the proposed configurations for each hyperparameter, and this will be executed for all five instruments. Q Learning and DQN will be tuned separately. Target weight update will be tuned for DDQN while copying the parameters for DQN.

\begin{table}[htbp]
  \centering
  \caption{Reinforcement Learning Hyperparameter Configurations for Grid Search}
    \begin{tabular}{lccccc}
          Hyperparameter               & Configurations                            \\
        \hline
          (All) Learning rate & 0.01, 0.001 \\ 
          (All) Discounted future reward factor & 0.9, 0.95 \\ 
          (DQN, DDQN) Hidden layer count and nodes & [32]*2, [64]*2  \\
          (DQN, DDQN) Target update frequency & 3, 6\\
          (DQN, DDQN) Batch size & 128, 256\\
          (DDQN) Target update weight & 0.1, 0.2\\
        \hline
    \end{tabular}
  \label{tab:reinforcementlearning_tuning}
\end{table}
There are four combinations for Q Learning, 32 combinations for DQN, and only two combinations for DDQN as it will share the parameters from DQN. It takes about 40 minutes per run, and given that there are five instruments to train on, this will take a total of about 125 hours (40*38*5/60).

\section{Evaluation}

This Section covers the results of the optimised models, evaluates them using performance metrics, and compares them to a benchmark (random action agent) using statistical significance testing. The models are tested through backtesting with historical held-out test data, and the best models are put through forward testing on the CTrader FXPro \cite{fxpro} platform. This Section concludes with an investigation to try to explain what values cause the agents to reverse their trades.

\subsection{Tuning Results}
The results from the hyperparameter tuning for the supervised learning and reinforcement learning components will be presented in this Section. This includes showing the best configurations for the parameters for each instrument, the respective learning curves during training, and the results of the held-out test data. Please note that the supervised learning and the reinforcement learning components are separate in this Section but will be combined for backtesting on the same test dataset in the next Section.

\subsubsection{Supervised Learning}
The supervised learning tuning methodology and table \ref{tab:supervisedlearningtuningresults} reveals the optimal parameters for each instrument.

\begin{table}[htbp]
  \centering
  \caption{Best Supervised Learning Hyperparameters from Tuning for Each Instrument}
    \begin{tabular}{lccccc}
          Parameter  & XAUUSD, GBPUSD,  EURUSD & FTSE100, DE40 \\
        \hline
          Layer/Nodes  & [2048]*4 &[2048]*4\\
           Learning Rate &\textbf{0.0001} & 0.00001\\
           Early Stopping Parameter &5&5\\
           Batch Size &256&256\\
        \hline
    \end{tabular}
  \label{tab:supervisedlearningtuningresults}
\end{table}
\newpage

It is reassuring to observe a lot of overlap in the parameters for each instrument because all of these networks are designed to do the same thing. The parameter distinguishing the models is the learning rate, which is ten times larger for XAUUSD, GBPUSD, and EURUSD compared to the indices FTSE100 and DE40. This suggests that XAUUSD, GBPUSD, and EURUSD are more noisy compared to the indices, which is supported by the high number of data points for the three instruments in table \ref{tab:ofiattributes}.

\begin{figure}[h!]
\centering
\includegraphics[width = \hsize]{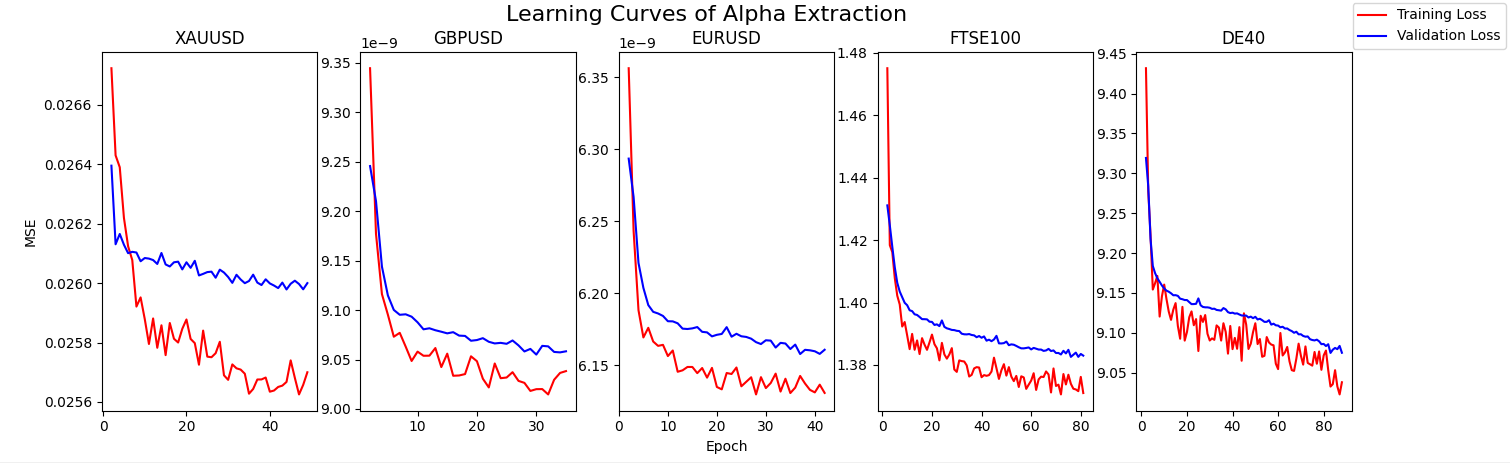}
\caption{Learning curves of alpha extraction showing training MSE loss (red) and validation loss (blue) for each instrument}
\label{fig:learningcurvesofalphaextraction}
\end{figure}

Figure \ref{fig:learningcurvesofalphaextraction} reveals the training and validation loss curves during training. The first few epochs are omitted because the losses were too large, and that reduced the detail toward termination. The training curve is quite noisy, but this is to be expected when dealing with market data. Overall, the curves are healthy because the validation loss is close to the training loss, but there are signs of overfitting when the gap is large, such as in XAUUSD (left-most panel). The other models also show overfitting but at a much smaller scale and may be acceptable.

\begin{table}[htbp]
  \centering
  \caption{Final Training, Validation, and Test Losses for Alpha Extraction (3 s.f.)}
    \begin{tabular}{lcccccc}
          Loss (MSE)  & XAUUSD & GBPUSD & EURUSD & FTSE100 & DE40 \\
          \hline
         Training  & 0.0257 & 9.04e-09 &  6.13e-09 & 1.37 & 9.04\\
          Validation  &0.0261& 9.06e-09&  6.16e-09& 1.38 & 9.08\\
          Test &  0.0249 &9.09e-09 &6.19e-06& 1.40 & 9.06 \\
          
    \end{tabular}
  \label{tab:mse}
\end{table}

Table \ref{tab:mse} presents the results at the end of training, and they are reasonable when compared with each other. For example, the models evaluated on the test set perform as well as on the validation set as expected, and they both are slightly above the training error. However, this does not reveal how good the models are. They need to be compared to a baseline. We use the results from Kolm et al. \cite{kolm} as our baseline, which requires the breaking down of the model's performance at the horizon level. This is done in table \ref{tab:rmseandsd}, and this presents the RMSE on the test set, the standard deviation of the test set, and the out-of-sample $R^2$ metric (calculated in equation \ref{eq:rsquared}) which is used by Kolm et al. \cite{kolm} to evaluate their work. There is a theme that the RMSE hovers just under the standard deviation, which is alarming as it indicates that there is a lot of overlap between the predictions and the rest of the data- ranging from 50\% to 67\% estimated overlap using \textsf{(2*(pnorm(q=RMSE/std,mean=0,sd=1) - 0.5))}. However, the average $R^2_{OS}$ performance matches that observed by Kolm et al. \cite{kolm} for alpha extraction using an MLP network. They observe an average of 0.181\% whereas the results in table \ref{tab:rmseandsd} show an average of 0.153\% and excluding the overfitted XAUUSD model gives an average of 0.180 so there is general confluence between these results.

\begin{table}[htbp]
  \centering
  \caption{Evaluating Alpha Extraction Error using out-of-sample $R^2_{OS}$ (\%) at each Horizon Level}
    \begin{tabular}{lcccccc}
          Attribute  & Horizon & XAUUSD & GBPUSD & EURUSD & FTSE100 & DE40 \\
          \hline
         Test Records & All & 523,799 & 358,196 & 272,335 & 187,851 & 171,758\\
        \hline
          RMSE &1 &0.0386& 2.14e-05& 1.85e-05& 0.348 & 0.92\\
          Std Dev. & 1 & 0.0313 &2.12e-05 &1.80e-05& 0.317 &1.02 \\
          R$^2_{OS}$ (\%) & 1 & -0.192 &-0.105  & -0.113& -0.080 & -0.056\\
          \hline
           RMSE &2& 0.0512& 3.04e-05& 2.52e-05& 0.405& 0.994\\
           Std Dev. & 2 & 0.0457& 3.12e-05 &2.60e-05 &0.431& 1.13\\
           R$^2_{OS}$ (\%) & 2 & -0.009 & 0.062 & 0.031&0.094 & 0.082\\
          \hline
           RMSE &3&  0.0626& 3.72e-05& 3.10e-05& 0.453& 1.18\\
           Std Dev. & 3& 0.0574& 3.92e-05& 3.23e-05 &0.527& 1.41\\
           R$^2_{OS}$ (\%) & 3 & 0.054 & 0.153 & 0.129& 0.186&0.173 \\
          \hline
           RMSE &4 & 0.0680& 4.17e-05& 3.38e-05& 0.502& 1.31\\
           Std Dev. & 4& 0.0673& 4.58e-05& 3.75e-05& 0.605 &1.54\\
           R$^2_{OS}$ (\%) & 4 & 0.078 &  0.199& 0.201& 0.304&0.367 \\
          \hline
           RMSE& 5&  0.0735& 4.64e-05& 3.78e-05& 0.554& 1.40\\
           Std Dev. & 5  &0.0758 &5.16e-05& 4.20e-05& 0.675 &1.73\\
           R$^2_{OS}$ (\%) & 5 & 0.152 &  0.267& 0.252 & 0.348& 0.391\\
          \hline
          RMSE &6 & 0.0827& 4.93e-05& 4.10e-05& 0.590& 1.47\\
          Std Dev. & 6 &0.0835 &5.67e-05 &4.61e-05& 0.737 &1.86\\
         R$^2_{OS}$ (\%) & 6 & 0.187  & 0.320 & 0.290&0.382 &0.428 \\
        \hline
        Average  \\
        R$^2_{OS}$ (\%) & All & \textbf{0.045} & \textbf{0.149} & \textbf{0.132} & \textbf{0.206} & \textbf{0.231}\\
        
        \hline
    \end{tabular}
  \label{tab:rmseandsd}
\end{table}

Kolm et al. \cite{kolm} also found that as the fraction of average price change increases (proportional to the horizon), the better the $R^2_{OS}$ (please see figure \ref{fig:kolmresults}) and table \ref{tab:rmseandsd} shows this for every instrument. This is perhaps because the higher the horizon, the higher the standard deviation, so the model is more likely to be relatively better compared to the benchmark. There is also agreement that the first horizon is worse than the benchmark used to calculate $R^2_{OS}$- negative $R^2$ value shown at horizon level 1. These confluences further support their results despite introducing their work to other asset classes, yet it is interesting to observe that the order in performance from best to worst is DE40, FTSE100, GBPUSD, EURUSD, and XAUUSD. The indices are outperforming the rest of the instruments, and Kolm et al. \cite{kolm} trained their models on stocks from the NASDAQ index, so equities might be ideal for this setup. Although these results agree with Kolm et al.'s outcomes, the $R^2_{OS}$ values infer that the models are able to explain 0.045\% to 0.231\% of the variance in alphas. Combining this alpha extraction with the reinforcement learning component will give more insight into the true performance of these supervised learning models.

\subsubsection{Reinforcement Learning}
The reinforcement learning tuning methodology (see tables \ref{tab:qlearnparams} and \ref{tab:dqnparams} below) shows the optimal parameters for each instrument for each agent. All three agents will be evaluated and compared together.

\begin{table}[htbp]
  \centering
  \caption{Best Q Learning Hyperparameters from Tuning for Each Instrument}
    \begin{tabular}{lccccc}
          Parameter  & XAUUSD & GBPUSD & EURUSD& FTSE100 & DE40 \\
        \hline
          Learning rate & 0.01 & 0.01 & 0.01 & 0.01  & 0.01\\
          Discounted future reward factor & \textbf{0.9} & 0.95& 0.95& 0.95& 0.95 \\
        \hline
    \end{tabular}
    
  \label{tab:qlearnparams}
\end{table}

\begin{table}[htbp]
  \centering
  \caption{Best DQN/DDQN Hyperparameters from Tuning for Each Instrument}
    \begin{tabular}{lccccc}
          Parameter  & XAUUSD & GBPUSD & EURUSD & FTSE100 & DE40 \\
        \hline
          Learning rate & 0.01 & 0.01 & 0.01 & 0.01 & 0.01 \\
          Discounted future reward factor & \textbf{0.9} & 0.95& 0.95& 0.95& 0.95 \\
          Layers/Nodes & [64]*2&[64]*2& [64]*2& [64]*2 & [64]*2  \\
          Target update frequency &  \textbf{6} & 3 &  3 & 3 & 3\\
          Batch size  &  128 &  128 &  128 &  128 &  128\\
          Target update weight & \textbf{0.1}& 0.2& 0.2& 0.2& 0.2\\
        \hline
    \end{tabular}
    
  \label{tab:dqnparams}
\end{table}

Once again, it is reassuring to see similar parameter values for each instrument, but there were not many options as shown in table \ref{tab:reinforcementlearning_tuning}. Similar to the supervised learning component, XAUUSD has a different configuration of parameters compared to the other instruments, perhaps due to the noise in its large dataset. This is confirmed by needing to update the target network less often and needing to update the target network parameters less to stabilise learning when compared with other instruments. The discounted future reward factor being lower also infers that alphas do not affect too far into the future due to noise.\\

\begin{figure}[h!]
\centering
\includegraphics[width = \hsize]{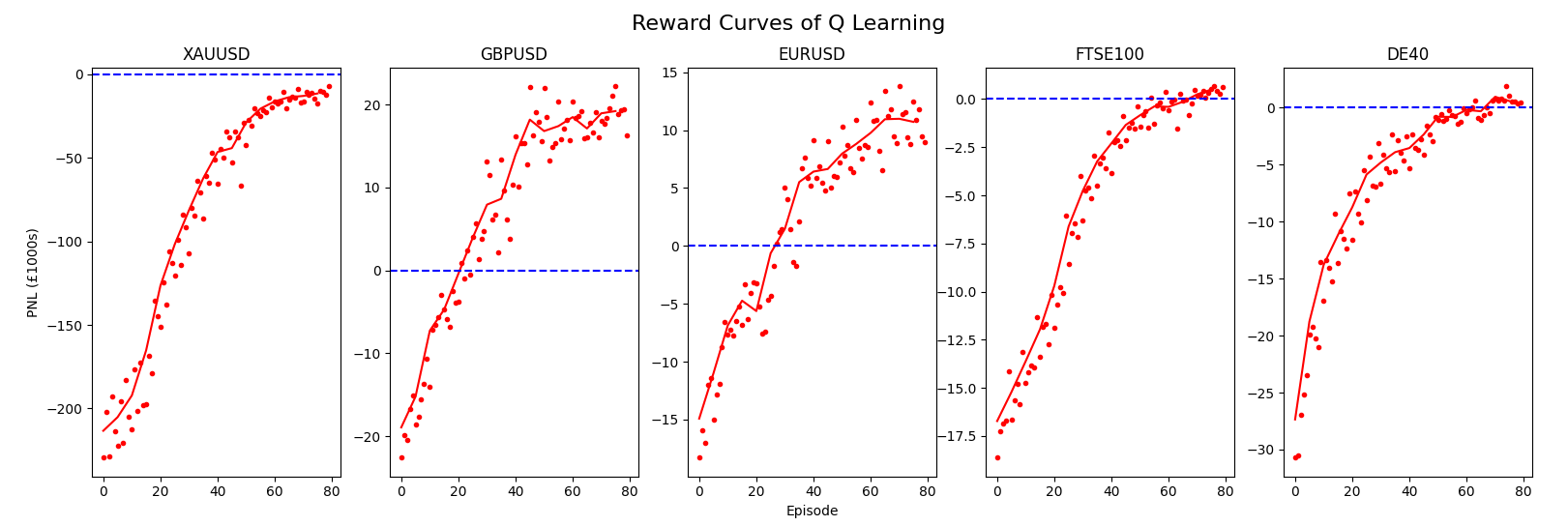}
\caption{Reward curves of Q Learning for each instrument during training (1 lot size trades)}
\label{fig:qlearningcurve}
\end{figure}

\begin{figure}[h!]
\centering
\includegraphics[width = \hsize]{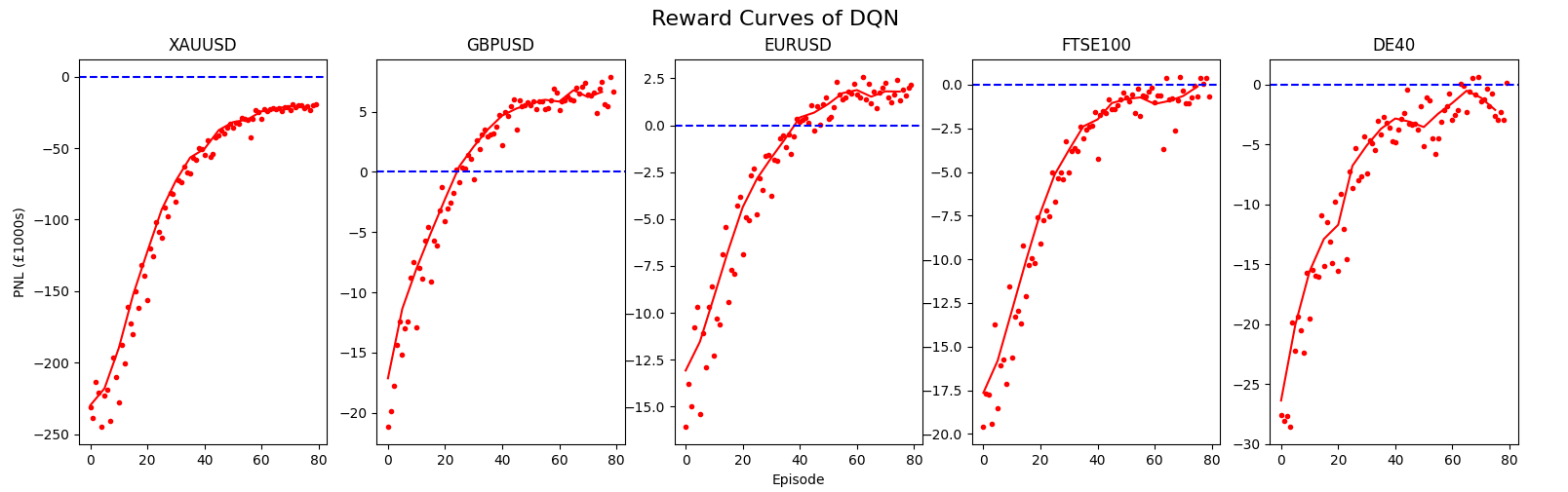}
\caption{Reward curves of DQN for each instrument during training (1 lot size trades)}
\label{fig:dqncurve}
\end{figure}

\begin{figure}[h!]
\centering
\includegraphics[width = \hsize]{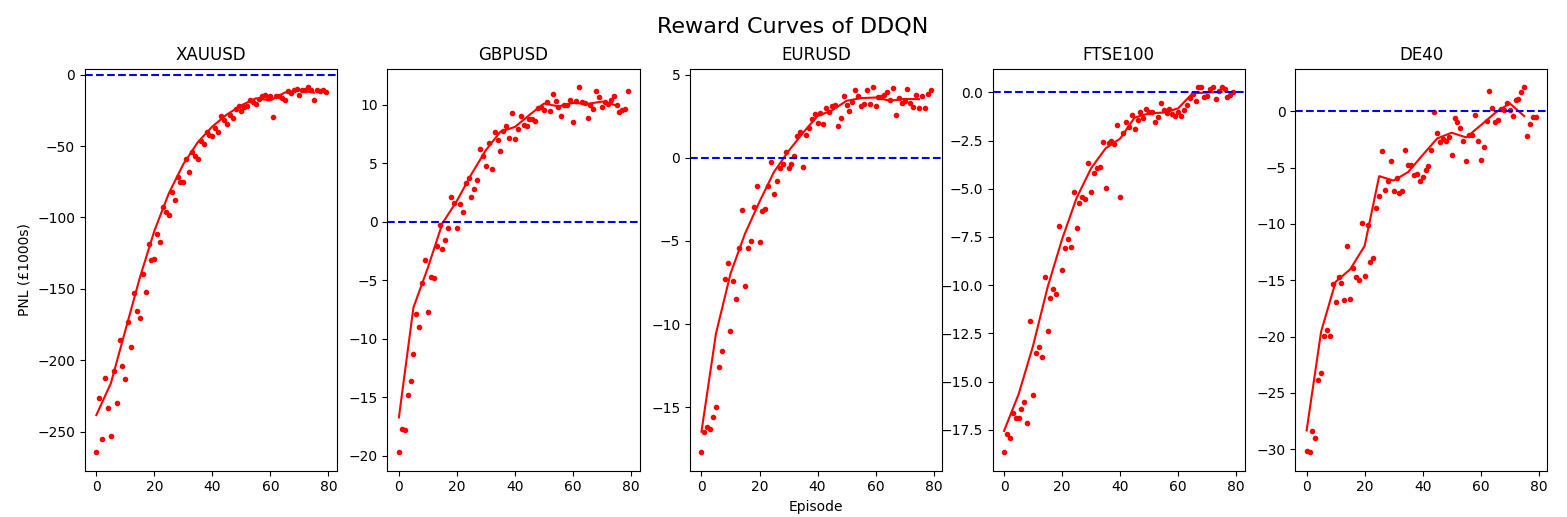}
\caption{Reward curves of DDQN for each instrument during training (1 lot size trades)}
\label{fig:ddqncurve}
\end{figure}

Figures \ref{fig:qlearningcurve}, \ref{fig:dqncurve}, and \ref{fig:ddqncurve} show the reward curves of each agent for each instrument during training. These curves overall indicate difficulty in learning, and this is because the agent needs to learn when it is worth reversing a trade, even though the alphas supplied are as accurate as they can be. The commissions applied to every trade match the costs at CTrader FXPro, which from lowest to highest per lot size is GBPUSD/EURUSD, XAUUSD, FTSE100, and DE40. It is surprising that FTSE100 and DE40 are close to being profitable during training across all agents, while XAUUSD was not profitable at all when having less transaction costs. This shows how poor the quality of the data collected for XAUUSD was given that it has almost all the other total records from the other instruments combined in the same ten weeks. With regards to the foreign exchange pairs having the least commissions, all the agents were very profitable, with Q Learning earning the highest, followed by DDQN, and then DQN- roughly with a ratio of 15:7:5. However, nothing is conclusive as this is only during training.\\

Tables \ref{tab:qtestingmetricsqlearning}, \ref{tab:dqntestingmetricsdqn}, and \ref{tab:ddqntestingmetricsddqn} show the calculated performance metrics of each agent on the 5-day test dataset for each instrument. Overall, the results align with the outcome of the reward curves during training, which indicates that these agents generalised well and there is little to no overfitting. Using the daily average profit and volatility, the order of best to worst instruments to apply these models to are GBPUSD, EURUSD, DE40, FTSE100, and XAUUSD.\\
Quantitatively, Q Learning made £8,030 in overall profit, whereas DQN lost £29,800 and DDQN lost £5,460, which is a wide range of outcomes. The row-wise average volatility (standard deviation) for the Q Learning agent (£1820) is almost double the average volatility for the DQN (£1060) and DDQN (£914), with DDQN having the smallest volatility. This can make the neural networks more suitable for stable trading once they become viable. The average profit/loss ratios and the profitability are very close together for DDQN and Q Learning (around 0.83,0.84 ratio with 49,51\% profitability) but noticeably worse for DQN (0.77 ratio with 44\% profitability).\\

The ranking of the agents from best to worst is Q Learning, DDQN, and DQN when it comes to overall execution. This supports Bertermann's \cite{bertermann} findings in that Q Learning is generally better than DQN for this environment. Although DDQN is more promising compared to DQN, as expected due to DQN's overestimation bias, Q Learning still outperforms DDQN. This indicates that tabular representation of the states is suitable compared to approximating the Q function as proposed by Bertermann \cite{bertermann} or perhaps more tuning is required to get the most out of the DQN/DDQN. However, it is important to note that Q tables have a limited domain range.

\begin{table}[htbp]
  \centering
  \caption{Q Learning Performance Metrics on Test Data (3 s.f.)}
    \begin{tabular}{lccccc}
          Metrics  & XAUUSD & GBPUSD & EURUSD & FTSE100 & DE40 \\
        \hline
          Daily Average Profit (£) & -20,100 & 18,600 & 9,890 & -252 &  -107\\
          Daily Average Volatility (£) & 3,390 & 3,280 & 1,810 & 274 & 363 \\
          Average Profit/Loss & 0.964 & 1.13 & 1.01 & 0.533 & 0.512 \\
          Profitability (\%)  & 35.9  & 62.8 & 62.3 & 43.3 & 52.4\\
          Maximum Drawdown (£) & -45,200 & -1.40 & -13.3 & -3,260 & -1,400\\
        \hline
    \end{tabular}
    
  \label{tab:qtestingmetricsqlearning}
\end{table}
\begin{table}[htbp]
  \centering
  \caption{DQN Performance Metrics on Test Data (3 s.f.)}
    \begin{tabular}{lccccc}
          Metrics  & XAUUSD & GBPUSD & EURUSD & FTSE100 & DE40 \\
        \hline
          Daily Average Profit (£) & -32,700 & 5,410 & 1,420 & -603 &  -3,300 \\
          Daily Average Volatility (£) & 1,360 & 1,560& 920& 503 & 946 \\
          Average Profit/Loss & 0.952 & 0.983 & 1.04 & 0.468 & 0.417  \\
          Profitability (\%)  & 20.1 & 53.5 & 54.2 & 51.6 & 42.9\\
          Maximum Drawdown (£) & -41,400 & -40.1 & -1,040& -2,500&-6,020\\
        \hline
    \end{tabular}
    
  \label{tab:dqntestingmetricsdqn}
\end{table}
\begin{table}[htbp]
  \centering
  \caption{DDQN Performance Metrics on Test Data (3 s.f.)}
    \begin{tabular}{lccccc}
          Metrics  & XAUUSD & GBPUSD & EURUSD & FTSE100 & DE40 \\
        \hline
          Daily Average Profit (£) & -16,900 & 8,090 & 3,410 & -55.4 &  -5.43\\
          Daily Average Volatility (£) & 1,430 & 1,380& 905& 103 & 751 \\
          Average Profit/Loss & 0.931 & 1.07 & 1.04 & 0.570 & 0.601  \\
          Profitability (\%)  & 38.5 & 57.3 & 57.9 & 44.2 & 51.7\\
          Maximum Drawdown (£) & -35,600 & -102& -1,100& -1,040 &-3,150\\
        \hline
    \end{tabular}
    
  \label{tab:ddqntestingmetricsddqn}
\end{table}

\subsection{Backtesting Results}

This Section combines both the alpha extraction (supervised learning component) and the reinforcement learning agents. OFI features will be taken as input to predict alphas using the supervised learning component, which will then be fed into the learning agents to output a trading signal. Tables \ref{tab:rltestingmetricsqlearningbt}, \ref{tab:rltestingmetricsdqnbt}, and \ref{tab:rltestingmetricsddqnbt} showcase the results of the pipeline on the same test data for comparison to tables \ref{tab:qtestingmetricsqlearning}-\ref{tab:ddqntestingmetricsddqn}.

\begin{table}[htbp]
  \centering
  \caption{Q Learning Performance Metrics on Test Data with Alpha Extraction (3 s.f.)}
    \begin{tabular}{lccccc}
          Metrics  & XAUUSD & GBPUSD & EURUSD & FTSE100 & DE40 \\
        \hline
          Daily Average Profit (£) & -126,000 & 2,210 & -120 & -4,100 & -2,990 \\
          Daily Average Volatility (£) & 9,150 & 3,590 & 2,100 & 593 & 675 \\
          Average Profit/Loss & 0.703 & 0.867& 0.841& 0.441 & 0.429 \\
          Profitability (\%)  & 23.6 & 53.6& 53.5 & 31.7& 38.3\\
          Maximum Drawdown (£) & -462,000 & -11,800& -9,510 & -15,100& -10,700\\
        \hline
    \end{tabular}
    
  \label{tab:rltestingmetricsqlearningbt}
\end{table}

\begin{table}[htbp]
  \centering
  \caption{DQN Performance Metrics on Test Data with Alpha Extraction(3 s.f.)}
    \begin{tabular}{lccccc}
          Metrics  & XAUUSD & GBPUSD & EURUSD & FTSE100 & DE40 \\
        \hline
          Daily Average Profit (£) & -151,000 & -11,500 & -12,000 & -4,830& -6,750 \\
          Daily Average Volatility (£) & 6,520 & 1,840& 1,530& 926 & 1,380 \\
          Average Profit/Loss & 0.684 &0.766 & 0.784& 0.410& 0.392 \\
          Profitability (\%)  &  19.1 & 43.6& 44.9 & 42.2& 39.6\\
          Maximum Drawdown (£) & -337,000 & -33,200& -38,000& -27,600&-29,900\\
        \hline
    \end{tabular}
    
  \label{tab:rltestingmetricsdqnbt}
\end{table}

\begin{table}[htbp]
  \centering
  \caption{DDQN Performance Metrics on Test Data with Alpha Extraction (3 s.f.)}
    \begin{tabular}{lccccc}
          Metrics  & XAUUSD & GBPUSD & EURUSD & FTSE100 & DE40 \\
        \hline
          Daily Average Profit (£) & -114,000 & -8,290 & -10,500 & -3,910& -3,010 \\
          Daily Average Volatility (£) & 6,780 & 2,150& 1,300& 491& 1,120 \\
          Average Profit/Loss & 0.719 & 0.855 &0.846 & 0.513& 0.549 \\
          Profitability (\%)  & 25.6 & 50.2& 51.4 &38.3 &45.9\\
          Maximum Drawdown (£) & -241,000 &-28,800 & -35,300& -29,100&-19,800\\
        \hline
    \end{tabular}
    
  \label{tab:rltestingmetricsddqnbt}
\end{table}
Overall, there is a noticeable drop in performance, which is expected because the alpha extraction is not the most accurate, while the learning agents were trained on accurate alphas. There is about a £100,000 decrease in daily average profit for gold, £15,000 decrease for the foreign exchange pairs, and £3,000 decrease for the indices- confirming the quality of the alpha extraction order regarding instruments already discussed. This brings all the profits from positive to negative except Q Learning on GBPUSD, which performs at a daily average of £2,210 with a standard deviation of £3,590. Q Learning on EURUSD follows in 2nd place at a daily average of -£120 with a standard deviation of £2,100, which shows potential as some days are profitable. The standard deviation has also increased in the range of 20\% to 200\%, increasing uncertainty as expected. The other metrics also indicate slightly poorer performance. All in all, Q Learning still outperforms DDQN, and DDQN outperforms DQN, matching the trend prior to incorporating the alpha extraction models.\\

Please note that the performance of these models might vary for another set of five days of test data, and so more data is required to make concrete conclusions about profitability. The analysis conducted in these sections should only be taken as an indication of performance.
\subsubsection{Benchmarking and Statistical Significance}

Although the backtesting results generally reveal poor performance apart from potential profitability with GBPUSD and EURUSD using Q Learning (further discussed in forward testing results), the outcome can be compared to the metrics of an agent that executes random actions on the same test dataset, and this is shown in table \ref{tab:rltestingmetricrandom}.

\begin{table}[htbp]
  \centering
  \caption{Random Action Performance Metrics on Test Data (3 s.f.)}
    \begin{tabular}{lccccc}
          Metrics  & XAUUSD & GBPUSD & EURUSD & FTSE100 & DE40 \\
        \hline
          Daily Average Profit (£)& -236,000  & -21,600  & -16,600 &-18,700 & -10,300  \\
          Daily Average Volatility (£)& 18,900  & 1,820 & 2,420 & 2,060& 1,200 \\
          Average Profit/Loss & 0.432 & 0.739 &  0.697 & 0.254 & 0.558 \\
          Profitability (\%)  & 18.5 & 43.2 & 42.8 & 12.6 & 40.1 \\
          Maximum Drawdown (£)  & -1,180,000& -108,000 &-82,900  & -93,500& -51,000 \\
        \hline
    \end{tabular}
    
  \label{tab:rltestingmetricrandom}
\end{table}

Executing on the same dataset allows for fair comparison, and statistical significance testing can be conducted to provide evidence, for formality, that the models perform better than the random agent. Due to there only being 5 data points (5 days of testing), it is better to use a non-parametric test such as the Mann-Whitney U-test (explained in \cite{whitney}). The daily profit from the trained agents and the random agent from backtesting on the test data will be used as input to the U-test. The one-tailed null hypothesis is that the trained models do not produce more profit compared to the random agent, and so the U value associated with each trained needs to be calculated and compared to the critical value in order to reject the null hypothesis.\\

The steps in \cite{whitney} were followed. All the sums of ranks for the Mann-Whitney U-test are 40, meaning that the profits from the trained models were all more than the profits from the random agent per day on the test dataset. Therefore, the process is the same for all the agents. The U value associated with each trained agent (for higher values) is 0. As ranks are always positive and the U value associated with each trained agent is 0, the null hypothesis is always rejected at all significant levels. Hence, the trained models are better than the random agent benchmark.

\subsection{Forward Testing Results}

This Section takes the agents performing well on certain instruments during backtesting and puts them in forward testing environments for real-time trading on the CTrader FXPro platform. As mentioned, this involves hosting a web application that will load the agents and wait for POST requests from the platform. The platform will send OFI data and retrieve an action, which will be executed on the platform.\\

\begin{figure}[h!]
\centering
\includegraphics[width =\hsize]{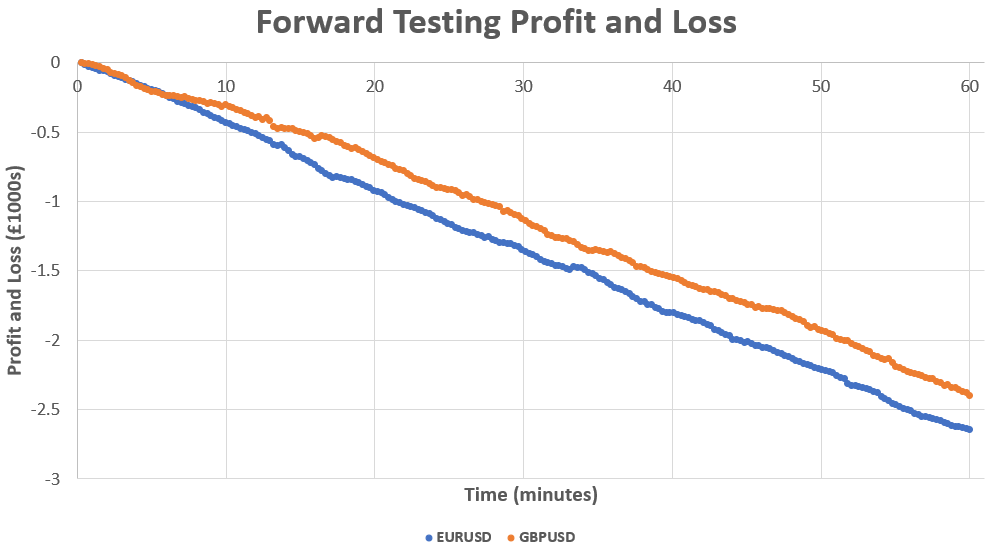}
\caption{Forward testing results using Q Learning on GBPUSD and EURUSD}
\label{fig:forwardtesting}
\end{figure}

The best-performing agents were selected based on the average daily profit being close to positive, and this was Q Learning on both GBPUSD and EURUSD. These will be the contenders to the forward testing pipeline. Forward testing was only conducted for an hour each, so due to the low sample size, the results can vary. Table \ref{fig:forwardtesting} shows the profit and loss graphs for Q Learning on GBPUSD and EURUSD.

Overall, it shows negative results with earning -£2,400 for GBPUSD and -£2,640 (£240 lower) for EURUSD, but there are a couple of reasons for this poor outcome. Firstly, the sample size is too small, and it could be profitable over longer periods of time, so more data is required to make a better judgement on its performance. This is less likely, though. Secondly and more importantly, the trading bot often skips a few timesteps while it processes the OFI values to return an action. During backtesting, the environment would wait for the agent to make an action before moving to the next step, whereas during forward testing, the environment does not wait and keeps moving, and then the trade is executed too late. Better hardware is required to mitigate this, and alternate solutions should be explored to replace the web application and have the agent be directly integrated with the platform to speed up the process. This was not possible from initial research.

\newpage
\subsection{Explainable AI}

\begin{figure}[h!]
\centering
\includegraphics[width =\hsize]{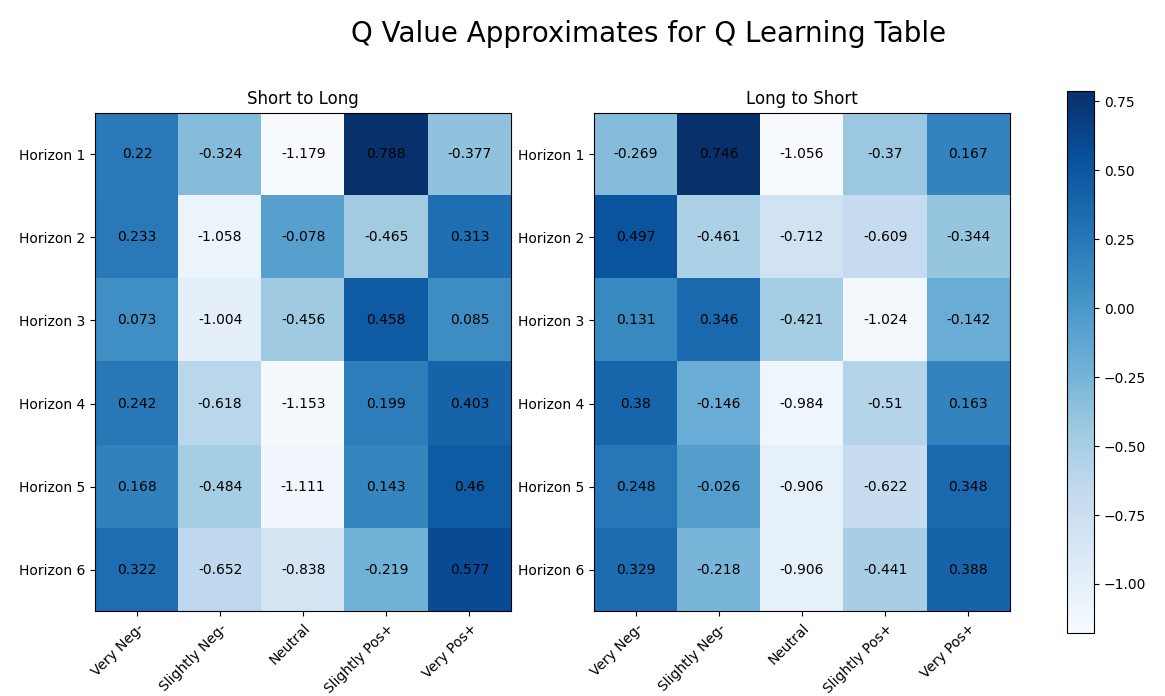}
\caption{Q value approximates for Q Learning table (averaged across all instruments)}
\label{fig:qexplained}
\end{figure}

\begin{figure}[h!]
\centering
\includegraphics[width =\hsize]{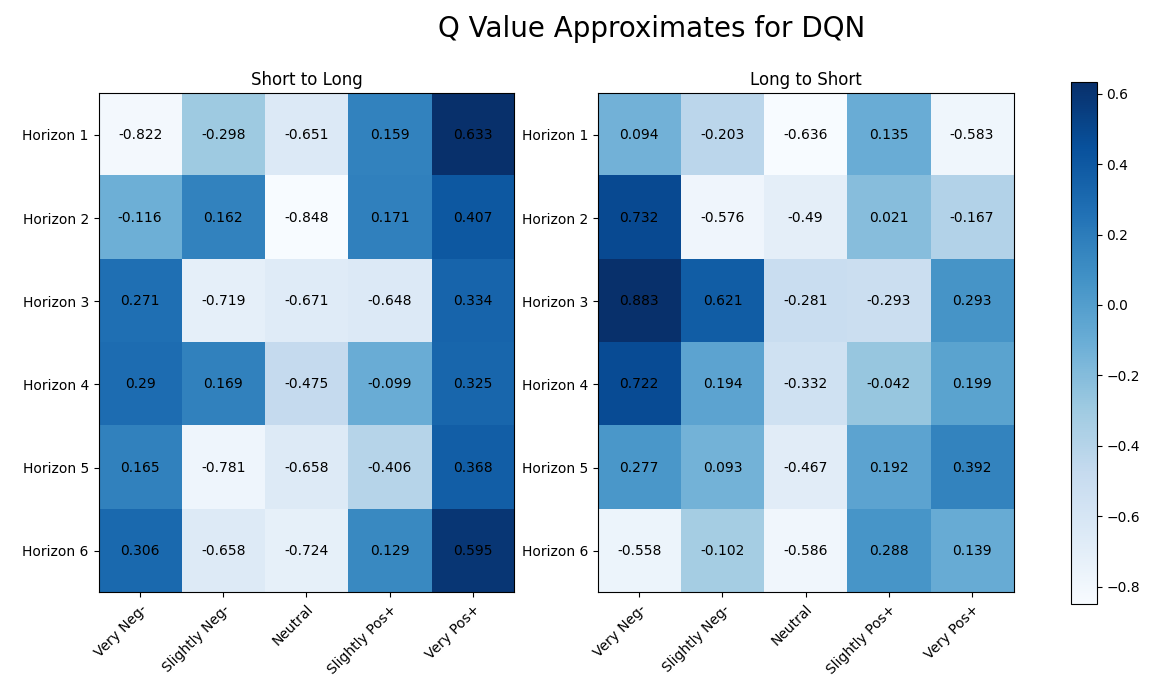}
\caption{Q value approximates for DQN (averaged across all instruments)}
\label{fig:dqnexpained}
\end{figure}

\begin{figure}[h!]
\centering
\includegraphics[width =\hsize]{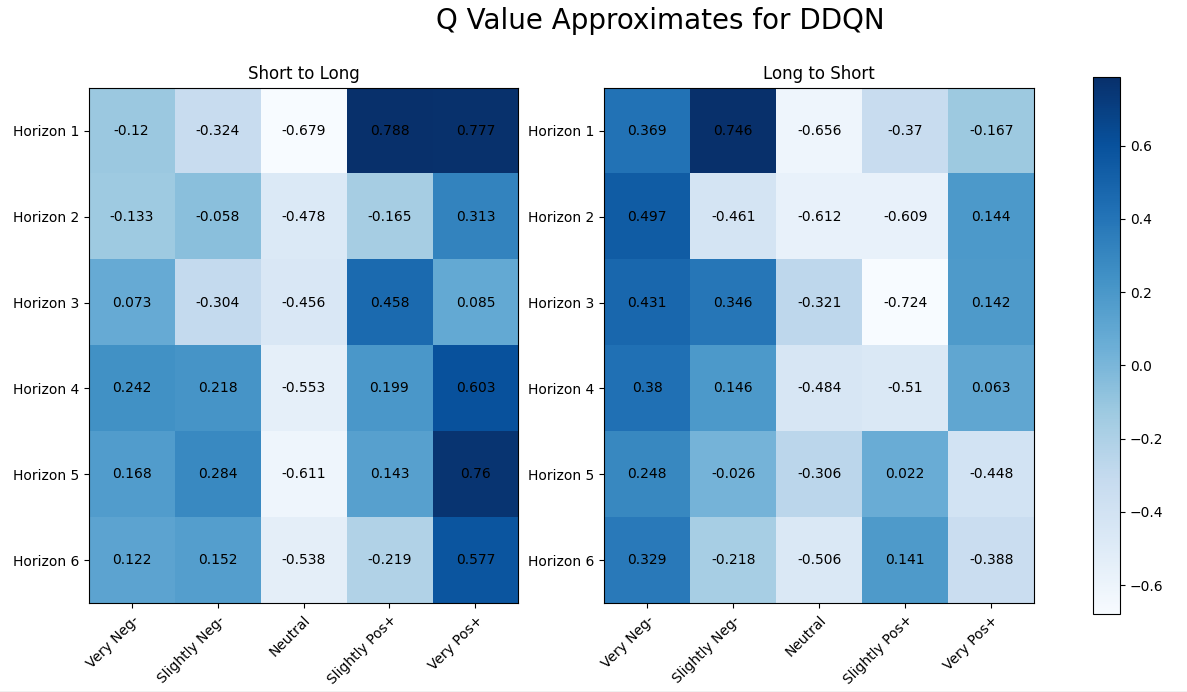}
\caption{Q value approximates for DDQN (averaged across all instruments)}
\label{fig:ddqnexplained}
\end{figure}

This Section attempts to explain the decisions made by the agents and investigates what horizon levels contribute the most to making the decision to send a reversal signal from short to long and short to long. Figures \ref{fig:qexplained}, \ref{fig:dqnexpained}, and \ref{fig:ddqnexplained} show heatmaps representing the average approximated Q values normalised across all instruments for each agent. DQN and DDQN are shown as tabular approximations by evaluating their functions at intervals equivalent to the bucket size for the Q Learning table to ensure a fair comparison. The state space (x-axis) contains 99.5\% of possible alpha values from the training dataset, which is 2.5 standard deviations from either side of the mean (close to 0).

Overall, these heatmaps show expected trends, such as having high positive Q values at positive alphas when choosing to go from short to long, and vice versa, with having high positive Q values at negative alphas when choosing to go from long to short. It is interesting to see that all agents have negative Q values in the neutral column, indicating that it is better not to reverse trades near alpha values of 0 (neutral), i.e. to avoid transaction costs when price will not move at all. This is expected, but the Q values are slightly positive at negative alphas when choosing to go from short to long and vice versa, which indicates a flaw in logic as it chooses not to trade at neutral alpha values at all. This is interesting because, in theory, it suggests that the agent is purposely trying to lose money, but in practice, the agent probably received enough reward whenever it did this, which paid off when the trade duration elapsed past the horizon window and worked out.\\

Although all the horizons are used to ultimately make the decision of whether to reverse the trade, the first horizon is predominantly used for Q Learning to both long and short, DQN to long, and DDQN to long and short, whereas the third horizon is used by DQN to short. The fifth and sixth horizons are used to long by DQN and DDQN, respectfully. However, all in all, the first horizon is mainly used to make decisions. This further supports why forward testing produced poor results, as the first horizon is almost always elapsed by the time a new trade was executed due to the relatively long process time of making predictions.
\newpage
\section{Conclusions and Further Work}

This research sets out to combine deep learning on the order books with reinforcement learning to break down large-scale end-to-end models into more manageable and lightweight components for reproducibility, suitable for retail trading. An alpha extraction model forecasts return over the next six timesteps, and a temporal-difference agent will use these values to provide trading signals (buy or sell). One alpha extraction model and three different temporal-difference agents were trained for five financial instruments, giving a total of 20 models and 15 trading bots.\\

The results for the different components align with the related literature. The alpha extraction outcome aligns with Kolm et al.'s \cite{kolm} results using the MLP architecture, and the rankings amongst the agents align with Bertermann's \cite{bertermann} findings. Only ten weeks of order flow imbalance data were collected for each instrument and split into 8:1:1 for training, validation, and testing. Grid search was used to find optimal parameters for each model, and more likely than not, values suggested in the literature were found to be best.\\

Overall, backtesting with retail costs produced promising results, with profitability achieved using Q Learning on GBPUSD and EURUSD, but failed to bring similar results during forward testing. This is due to long processing times of making predictions, sometimes skipping two timesteps, but this can be mitigated with a different infrastructure to using web applications as well as with more advanced hardware. The current setup uses Intel i9-9900K CPU @ 3.60GHz 16GB and Nvidia GeForce RTX 2080 Super 16GB.\\

The agents, on top of learning expected patterns, also learned patterns that were undesirable due to exposure to receiving positive rewards beyond the horizon window despite observing contradictory forecasting values. This can be solved by increasing this horizon window to beyond six. Other improvements consist of collecting more data, using better hardware, using rate of return instead to standardise across multiple instruments, using another trading platform with lower commissions, and expanding the action space to also not be in any position (i.e. three actions of buy, sell, and not be in a position instead of just buy and sell). \\

With regard to legal, social, ethical, and professional considerations, the only concern is scraping raw limit order book data from the CTrader platform and storing it locally, which raises ethical issues. Scraping is not a legal violation of their EULA \cite{eula} as this work is for research and not commercial gain. This issue was minimised by only storing wanted order flow imbalance features inferred from the raw limit order book states instead of storing actual raw states directly.


\appendix
\section{Supplementary Figures}

\begin{figure}[h!]
\centering
\includegraphics[width = \hsize]{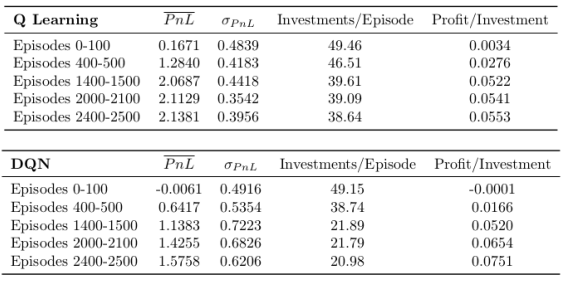}
\caption{Tabular performance comparison of Q Learning and DQN provided by Bertermann \cite[Page 29]{bertermann}}
\label{fig:bertermannresults}
\end{figure}

\begin{table}[htbp]
  \centering
  \caption{Mean of Values at each OFI Level from the Collected Data (4 d.p.)}
    \begin{tabular}{lccccc}
          OFI Level  & XAUUSD & GBPUSD & EURUSD & FTSE100 & DE40 \\
        \hline
           1  & 0.1236 &0.0054& 0.0015& 0.0251& 0.0350\\
           2   &0.1032 &0.0312& 0.0303& 0.1119 &0.0073\\
           3   &0.1241 &0.0567& 0.0569& 0.1299 &0.0122\\
           4  & 0.1488 &0.1037& 0.1016& 0.1334 &0.0175\\
           5   &0.1763 &0.1470 &0.1510& 0.1780& 0.0226\\
          \hline
           6   &0.2043 & 0.1679& 0.1313& 0.2077 &0.0564 \\
           7   &0.0566 & 0.0875 &0.0489& 0.1057 & 0.0113\\
           8   & 0.0779& 0.1190 &0.0913& 0.0818& 0.0097\\
           9   &0.0956 &0.0828 & 0.1341 &0.0456& 0.0143 \\
           10  & 0.1078& 0.0445 &0.0282& 0.0332 &0.0163 \\
        \hline
    \end{tabular}
  \label{tab:ofimean}
\end{table}

\newpage

\begin{table}[htbp]
  \centering
  \caption{Standard deviation of Values at each OFI Level from the Collected Data (4 d.p.)}
    \begin{tabular}{lccccc}
          OFI Level  & XAUUSD & GBPUSD & EURUSD & FTSE100 & DE40 \\
        \hline
           1 & 0.2814 &0.1267& 0.0774 &0.3985& 0.9989 \\
           2  & 0.3444 &0.2261 &0.2730 & 0.6038& 0.1844\\
           3  & 0.3962 &0.3301 &0.3995& 0.4512& 0.1985\\
           4  &0.4553& 0.4233& 0.4646& 0.4819& 0.2021\\
           5 & 0.5133 &0.4910 & 0.4674& 0.5603 &0.2149\\
        \hline
           6  &0.5674 &0.5434& 0.4190 & 0.4896& 0.2553\\
           7 & 0.3273 &0.4199& 0.3591& 0.4857& 0.1846\\
           8 & 0.3775& 0.4572& 0.4729& 0.5083& 0.2177\\
           9 & 0.4016 &0.3837 &0.5153 &0.4034 &0.2152\\
           10  &0.4302 &0.4472& 0.4673& 0.3846& 0.1892\\
        \hline
    \end{tabular}
  \label{tab:ofistd}
\end{table}

\begin{table}[htbp]
  \centering
  \caption{Percentage of Positive Entries at each OFI Level from the Collected Data (1 d.p.)}
    \begin{tabular}{lccccc}
          OFI Level  & XAUUSD & GBPUSD & EURUSD & FTSE100 & DE40 \\
        \hline
           1 & 52.8& 53.9& 53.3 &53.1& 51.7 \\
           2  &  66.1 &62.2& 60.5& 55.8& 53.1\\
           3  & 67.0&  62.8& 60.2& 53.5& 57.5\\
           4  &67.2& 64.0 & 61.9 &49.1 &56.9 \\
           5 & 67.8& 64.5& 65.2& 46.6 &56.2\\
        \hline
           6  &67.8 &62.6& 61.7& 35.3& 52.7\\
           7 & 25.1 &35.6 &36.7 &24.6& 21.9\\
           8 & 24.3 &31.2 &35.4 &21.6& 21.0\\
           9 & 22.5 &21.2 &30.1 &19.2 &20.9\\
           10  &18.9 &15.3 &15.0 & 17.5& 21.2\\
        \hline
    \end{tabular}
  \label{tab:ofipos}
\end{table}

\begin{table}[htbp]
  \centering
  \caption{Percentage of Negative Entries at each OFI Level from the Collected Data (1 d.p.)}
    \begin{tabular}{lccccc}
          OFI Level  & XAUUSD & GBPUSD & EURUSD & FTSE100 & DE40 \\
        \hline
           1 & 19.6 &32.3& 34.5& 46.9& 48.2\\
           2  & 28.2 &37.0 & 39.2& 43.0&  46.6\\
           3  &28.2& 36.4 &39.6& 29.6 &42.1\\
           4  &28.9& 34.2 &37.5& 24.2 &41.2 \\
           5 & 29.3 &32.7& 33.4& 23.3 &40.4\\
        \hline
           6  &29.4& 32.0 & 29.2 &10.&  32.1\\
           7 &12.3 &19.6& 22.8 &12.4& 18.9\\
           8 & 11.6& 15.5& 20.9 &14.9& 16.9\\
           9 & 9.8 &10.6& 15.5 &15.4& 17.7\\
           10  &7.1 & 9.3& 10.8& 15.0&  16.7\\
        \hline
    \end{tabular}
  \label{tab:ofineg}
\end{table}

\newpage

\begin{table}[htbp]
  \centering
  \caption{Percentage of Zero Entries at each OFI Level from the Collected Data (1 d.p.)}
    \begin{tabular}{lccccc}
          OFI Level  & XAUUSD & GBPUSD & EURUSD & FTSE100 & DE40 \\
        \hline
           1 & 27.6& 13.8& 12.2&  0.0 &  0.1 \\
           2  &5.7 & 0.9 & 0.3 & 1.1 & 0.3\\
           3  &4.8 & 0.9 & 0.2 &16.8  &0.4\\
           4  & 3.9 & 1.8 & 0.7& 26.6 & 1.9 \\
           5 & 2.9&  2.9  &1.4& 30.1 & 3.4\\
        \hline
           6  &2.7&  5.4 & 9.1& 54.7 &15.2\\
           7 &62.6 &44.8 &40.5 &63.0 & 59.3\\
           8 & 64.1 &53.2& 43.6& 63.5& 62.1\\
           9 & 67.7 &68.2 &54.4 &65.4& 61.3\\
           10  &74.0&  75.3 &74.2 &67.5 &62.1\\
        \hline
    \end{tabular}
  \label{tab:ofizer}
\end{table}

\begin{figure}[h!]
\centering
\includegraphics[width = 0.9\hsize]{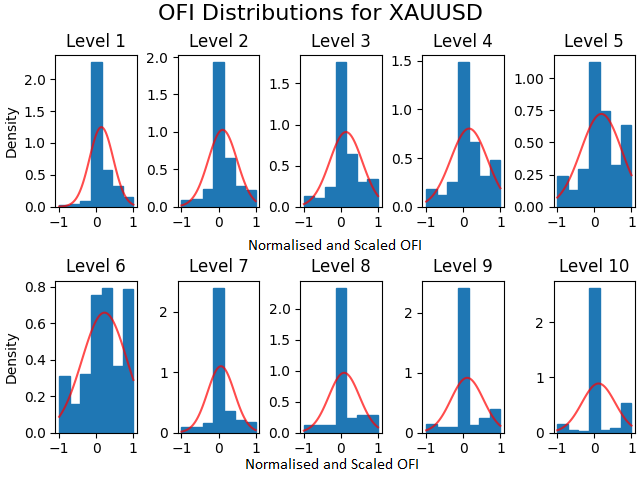}
\caption{Distribution of first ten OFI levels (blue) from collected XAUUSD (gold) data against a fitted normal distribution (red)}
\label{fig:ofidist_xauusd}
\end{figure}

\newpage

\begin{figure}[h!]
\centering
\includegraphics[width = 0.75\hsize]{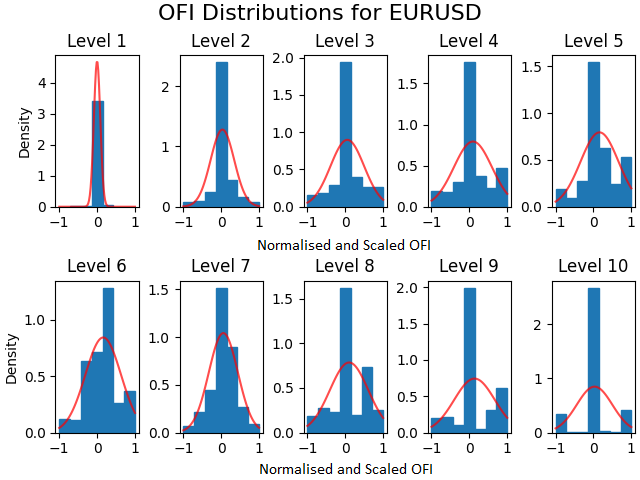}
\caption{Distribution of first ten OFI levels (blue) from collected EURUSD data against a fitted normal distribution (red)}
\label{fig:ofidist_eurusd}
\end{figure}

\begin{figure}[h!]
\centering
\includegraphics[width = 0.75\hsize]{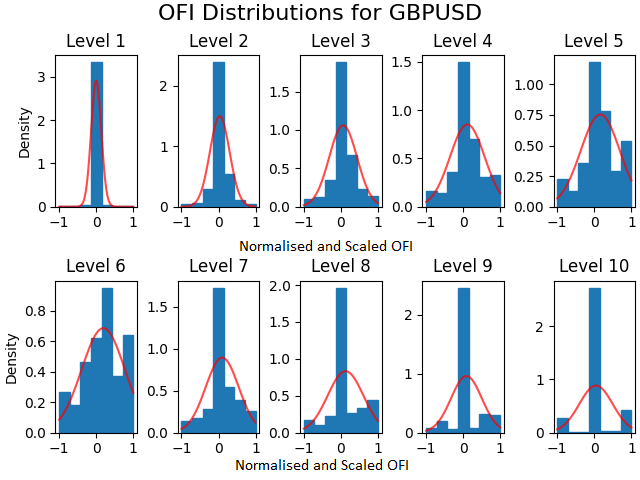}
\caption{Distribution of first ten OFI levels (blue) from collected GBPUSD data against a fitted normal distribution (red)}
\label{fig:ofidist_gbpusd}
\end{figure}

\newpage

\begin{figure}[h!]
\centering
\includegraphics[width = 0.75\hsize]{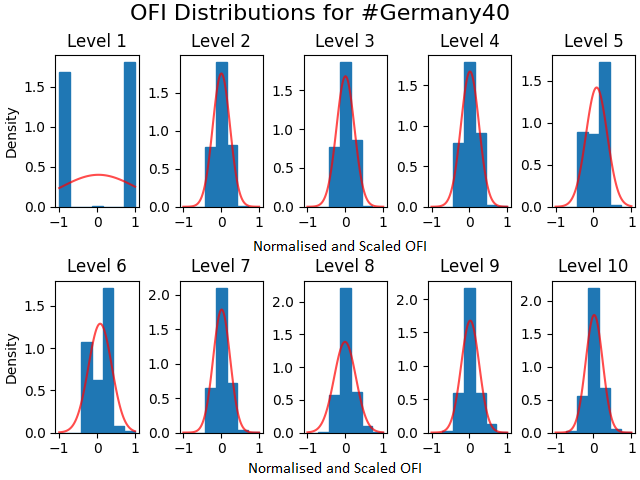}
\caption{Distribution of first ten OFI levels (blue) from collected DE40 data against a fitted normal distribution (red)}
\label{fig:ofidist_de40}
\end{figure}

\begin{figure}[h!]
\centering
\includegraphics[width = 0.75\hsize]{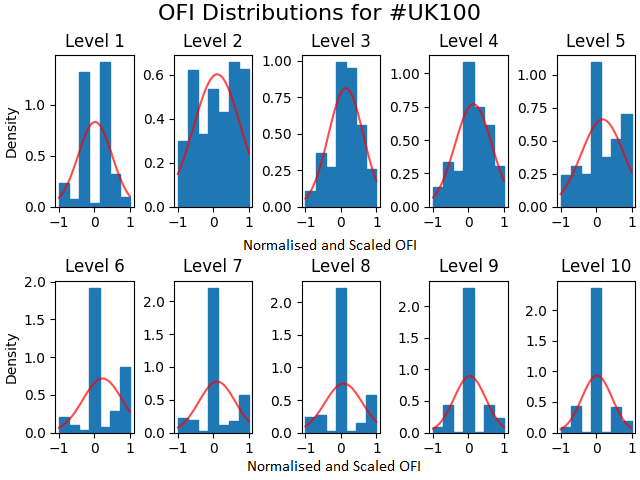}
\caption{Distribution of first ten OFI levels (blue) from collected FTSE100 data against a fitted normal distribution (red)}
\label{fig:ofidist_ftse100}
\end{figure}

\newpage

\begin{table}[htbp]
  \centering
  \caption{Mean of Alphas (in Pips) at each Horizon from Collected Data (3 s.f.)}
    \begin{tabular}{lccccc}
          Horizon  & XAUUSD & GBPUSD & EURUSD & FTSE100 & DE40 \\
        \hline
           1  & -8.11e-04 & 3.37e-05 &-1.12e-05& -4.09e-03& -5.07e-03\\
           2   &-1.63e-03 & 6.69e-05 &-2.35e-05 &-8.15e-03& -1.02e-02\\
           3   &-2.44e-03 & 9.96e-05 &-3.62e-05& -1.22e-02& -1.53e-02\\
          \hline
           4  &-3.26e-03 & 1.31e-04& -4.87e-05 &-1.63e-02 &-2.05e-02\\
           5   &-4.08e-03 & 1.63e-04 &-6.14e-05& -2.03e-02 &-2.56e-02\\
           6   &-4.90e-03 & 1.95e-04 &-7.40e-05 &-2.44e-02& -3.07e-02 \\
        \hline
    \end{tabular}
  \label{tab:alphamean}
\end{table}

\begin{table}[htbp]
  \centering
  \caption{Standard Deviation of Alphas (in Pips) at each Horizon from Collected Data (3 s.f.)}
    \begin{tabular}{lccccc}
          Horizon  & XAUUSD & GBPUSD & EURUSD & FTSE100 & DE40 \\
        \hline
           1  & 2.36& 0.155& 0.138& 3.10& 10.1\\
           2   &3.48& 0.230& 0.200& 4.08& 10.7\\
           3   &4.38& 0.289& 0.251 &4.94& 13.5\\
          \hline
           4  & 5.14& 0.339& 0.293& 5.64& 14.6\\
           5   &5.81& 0.383 &0.330& 6.28& 16.4\\
           6   &6.41 &0.423& 0.363& 6.84& 17.5 \\
        \hline
    \end{tabular}
  \label{tab:alphastd}
\end{table}
\begin{table}[htbp]
  \centering
  \caption{Percentage of Positive Alphas at each Horizon from Collected Data (1 d.p.)}
    \begin{tabular}{lccccc}
          Horizon  & XAUUSD & GBPUSD & EURUSD & FTSE100 & DE40 \\
        \hline
           1  & 28.8 &41.3 &39.9& 46.6 &48.2\\
           2   &32.7& 30.0 & 28.7& 28.3 &16.3\\
           3   &38.0 & 41.5& 39.8 &44.4& 47.8\\
          \hline
           4  & 39.8& 36.9& 35.5 &35.0 & 23.5\\
           5   &41.8 &42.4 &40.8 &44.2& 47.5\\
           6   &42.8&40.0 & 38.8 &38.6& 28.0 \\
        \hline
    \end{tabular}
  \label{tab:alphapos}
\end{table}
\begin{table}[htbp]
  \centering
  \caption{Percentage of Negative Alphas at each Horizon from Collected Data (1 d.p.)}
    \begin{tabular}{lccccc}
          Horizon  & XAUUSD & GBPUSD & EURUSD & FTSE100 & DE40 \\
        \hline
           1  & 29.4& 41.2& 39.8& 46.9 &48.2\\
           2   &33.2& 30.0 & 28.8 &28.2 &16.4\\
           3   &38.6& 41.5& 39.9& 44.4& 47.7\\
          \hline
           4  & 40.4 &36.8& 35.7& 35.0 & 23.5\\
           5   &42.5& 42.3 &41.0 & 44.1& 47.4\\
           6   &43.4& 40.0 & 39.0 & 38.5 &28.0 \\
        \hline
    \end{tabular}
  \label{tab:alphaneg}
\end{table}
\begin{table}[htbp]
  \centering
  \caption{Percentage of Zero Alphas at each Horizon from Collected Data (1 d.p.)}
    \begin{tabular}{lccccc}
          Horizon  & XAUUSD & GBPUSD & EURUSD & FTSE100 & DE40 \\
        \hline
           1  & 41.8 &17.5 &20.3 & 6.5  &3.6\\
           2   &34.1 &40.0 & 42.5& 43.5 &67.3\\
           3   &23.4& 17.0 & 20.3& 11.2  &4.5\\
          \hline
           4  & 19.8& 26.3& 28.8 &30.0 & 53.0 \\
           5   &15.7 &15.3 &18.2& 11.7 & 5.1\\
           6   &13.8 &20.0 & 22.2 &22.9 &44.0 \\
        \hline
    \end{tabular}
  \label{tab:alphazer}
\end{table}

\begin{figure}[h!]
\centering
\includegraphics[width = \hsize]{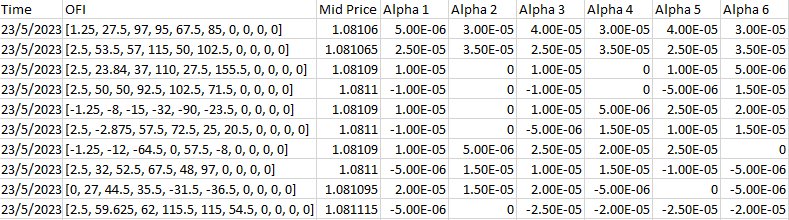}
\caption{First ten records of the CSV file containing the alpha data calculated from the collected data}
\label{fig:alphas}
\end{figure}

\end{document}